\documentclass[10pt,letterpaper,twocolumn]{article}
\usepackage[english]{babel}
\usepackage{graphicx}

\newcommand{\alt}{\mathbin{\lower 3pt\hbox
   {$\rlap{\raise 5pt\hbox{$\char'074$}}\mathchar"7218$}}}
\newcommand{\agt}{\mathbin{\lower 3pt\hbox
   {$\rlap{\raise 5pt\hbox{$\char'076$}}\mathchar"7218$}}}

\begin{document}

\setcounter{footnote}{0}
\setcounter{equation}{0}
\setcounter{figure}{0}
\setcounter{table}{0}

\title{\large\bf Conductance distribution in 1D
systems: \\
dependence on the Fermi level and the ideal leads }

\author{\small I. M. Suslov  \\
\small P.L.Kapitza Institute for Physical Problems,  \\
\small 119334 Moscow, Russia  \\
\small E-mail: suslov@kapitza.ras.ru\\
 {}\\
\parbox{120mm}{\footnotesize \,The correct definition
of the conductance of finite systems implies a
connection to the system of the massive ideal leads. Influence
of the latter on the properties of the system appears to be
rather essential and is studied below on the simplest example
of the 1D case. In the log-normal regime this influence is
reduced to the change of the absolute scale of conductance, but
generally changes the whole distribution function. Under the
change of the system length $L$, its resistance may
undergo the periodic or aperiodic oscillations. Variation
of the Fermi level induces  qualitative changes in the
conductance distribution, resembling the smoothed Anderson
transition. } }

\date{}
\maketitle

\textwidth 6.4 in
\textheight 8.5 in

\setcounter{footnote}{0}
\setcounter{equation}{0}
\setcounter{figure}{0}
\setcounter{table}{0}

\begin{center}
{\bf 1. Introduction and main results}
\end{center}

The correct definition of the conductance of finite systems
is not a trivial issue, which was a subject of the vivid
discussion at early 1980-ties \cite{1}--\cite{9} (see the
review article \cite{10}). The reason of controversy was related
to a fact that the conductance of finite systems is an
ill-defined quantity. It is a consequence of the specific
feature of the linear response formulas: the $\delta$-functions,
contained in them,  should be extended to the width $\Gamma$,
which should be tended to zero only after the thermodynamic limit
transition; such procedure is surely impossible in finite
systems. To avoid this difficulty, the rather elegant trick was
suggested \cite{3}:  the finite system is connected to
the ideal leads (Fig.1), which are suggested to be sufficiently
massive, so the thermodynamic limit is practically taken in these
leads. Such construction solves the problem of interpretation
of the Kubo formula  but creates  new problems: the
corresponding definition of conductance refers to the
composite system \makebox{"sample+ideal leads",} and its relation
to the initial system remains the open question. In order to
clarify a situation, one can introduce the semi-transparent
interfaces between the system and the ideal leads \cite{11}.
Influence of the latter on the properties of the system is surely
essential near the Anderson transition \cite{12}, but was
discussed in \cite{11,12} only on the abstract level. In fact,
this influence can be a subject of the constructive analysis,
which is demonstrated below on the simplest example of 1D
systems.

In theoretical papers, one usually suggests the existence of the
unique Fermi level, while a difference between the sample
and ideal leads
is determined by the absence of the random potential in the latter
case (Fig.2,a).  Practically, such situation is not very
realistic: usually, in the experimental device the connecting
leads are produced  from a good metal with the large Fermi
energy (so impurities are effectively screened), while
the typical disordered system is a semi-metal or a doped
semiconductor, where effects of disorder
are strikingly  manifested  (Fig.2,b).
Nevertheless, the  difference from the former case (Fig.2,a) does
not look essential and is usually ignored in theoretical papers.
However, the use of the foreign leads becomes inevitable,
if the Fermi level in the system corresponds to the
forbidden band of the ideal crystal (Fig.2,c), where allowed
states arise
only due to a random potential. In this case, the attempt to
make the connecting leads from the same material without
impurities will end by a confusion: "the ideal leads" will
not conduct at all.

An explicit introduction of the ideal leads (Fig.2,\,b,\,c) is
realized with a help of the edge transfer matrices (Sec.2)
and sets the interesting physical problem:
producing  small variation of the Fermi level
(not essential
for the ideal leads), one can transfer from the
quasi-metallic regime in the allowed band (Fig.2,b) to the
fluctuational states in the deep of the forbidden band
(Fig.2,c). In the course of it, the conductance distribution
of the 1D system undergoes the qualitative changes resembling the
smoothed Anderson transition.

If $g$ is the dimensionless conductance (conductance $G$ of the
system of size $L$ in quantum units  $e^2/h$), then the
distribution $P(\rho)$ of dimensionless resistance $\rho=1/g$ is
described by the following evolution equation
$$
\frac{\partial P(\rho)}{\partial L} =
\alpha\,\frac{\partial}{\partial \rho}
\left[\,\rho(1\!+\!\rho)\,\frac{\partial P(\rho)}{\partial \rho}
\,\right] \,, \eqno(1)
$$
derived in the number of papers \cite{13}--\cite{18} and
considered as sufficiently universal. However, more general
equation was suggested in \cite{12} for 1D systems
$$
\frac{\partial P(\rho)}{\partial L} =
\tilde\alpha\,\frac{\partial}{\partial \rho}
\left[\,-\gamma(1\!+\!2\rho) P(\rho) +
\rho(1\!+\!\rho)\,\frac{\partial P(\rho)}{\partial \rho}
\,\right]   \,,
\eqno(2)
$$
which is reduced to (1) in the random phase approximation.
The latter approximation is sufficiently good for the deep
of the allowed band and the "natural" ideal leads (Fig.2,a),
which is usually suggested in theoretical papers (see
references in \cite{19,20}), while the situation in the
forbidden band is considered infrequently \cite{21,22,23}
and only on the level of wave functions.
The main argument of the paper \cite{12} was based on the fact,
that a finite value of the parameter  $\gamma$ arises in the case
of semi-transparent boundaries (Fig.1), even if the random phase
approximation is applicable to the system under consideration.

\begin{figure}[t]
\centerline{\includegraphics[width=2.6 in]{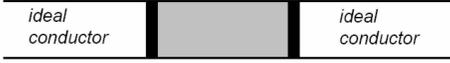}}
\caption{\footnotesize The correct definition of the conductance
of finite systems implies a connection of the massive ideal
leads. For the discussion of influence of the latter,
one can introduce the semi-transparent boundaries
\cite{11}.  }
\label{fig1} \end{figure}
\begin{figure}
\centerline{\includegraphics[width=2.5 in]{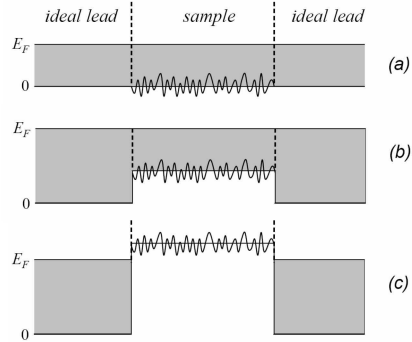}} \caption{
\footnotesize
(a) In the theoretical papers one usually suggests
that a difference between the sample and  ideal leads
is determined by the absence of a random potential in the latter
case; (b) Practically in the
experiment the  connecting leads are produced from a good metal
with the large Fermi energy;  (c) Explicit introduction of the
ideal leads is inevitable, if the Fermi level corresponds
to the forbidden band of the ideal crystal.
} \label{fig2}
\end{figure}

Still more general evolution equation arises for the explicitly
introduced ideal leads (Sec.7)
$$
\frac{\partial P(\rho)}{\partial L} =
\tilde\alpha\,\frac{\partial}{\partial \rho}
\left[\,-\gamma_1(1\!+\!2\rho) P(\rho)
\vphantom{\frac{\partial P(\rho)}{\partial \rho}}
\,-\right.   \qquad\qquad
$$
$$
\qquad\qquad\left.
-2\gamma_2 \sqrt{\rho(1\!+\!\rho)} P(\rho)
+\rho(1\!+\!\rho)\,\frac{\partial P(\rho)}{\partial \rho}
\,\right]   \,,
\eqno(3)
$$
which is reduced to (2) in the regions of small and large $L$,
where the typical values of $\rho$ are small and large
correspondingly (then $\gamma=\gamma_1$  in the first case
and $\gamma=\gamma_1+\gamma_2$ in the second one), so equation
(3) is analogous to equation (2) with variable $\gamma$.
A possibility of the $\gamma$ variation  in the course of
evolution was allowed in \cite{12} from the very beginning
and is systematically studied in Sec.3. The limiting
value of $\gamma$ for large $L$ is determined by
internal properties of the system under consideration
and does not depend on ideal leads. Its behavior as function
of ${\cal E}/W^{4/3}$ is shown in Fig.3, where ${\cal E}$
is the Fermi energy measured from the lower band edge
and  $W$ is the amplitude of a random potential (all energies
are measured in units of the hopping integral of the 1D Anderson
model, see Eq.14 below). One can see that the parameter
$\gamma$ is always finite but accepts small values in
the deep of the allowed band, in accordance with the random phase
approximation. Hence, finiteness of $\gamma$ is determined by
the internal properties of the system and introduction of
semi-transparent boundaries  \cite{12} is not actual.

\begin{figure}
\centerline{\includegraphics[width=2.7 in]{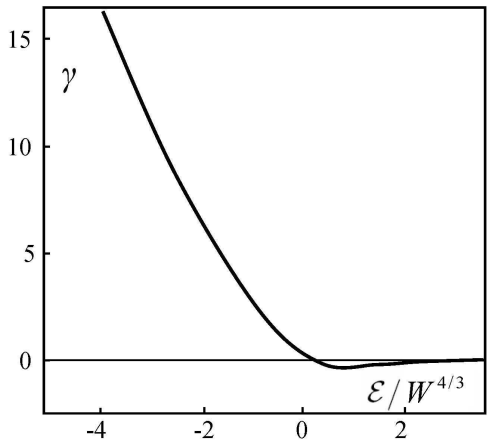}}
\caption{\footnotesize Parameter $\gamma$ in equation (2),
corresponding to the limit of large $L$, as function
 of the energy ${\cal E}$,
counted from the lower edge of the initial band.
} \label{fig3}
\end{figure}

In the plane $({\cal E},W^2)$ one can distinguish three
characteristic domains (Fig.4): quasi-metallic ($|\gamma|\ll 1$),
strongly localized ($\gamma\gg 1$) and "critical" ($\gamma\sim
1$). If the energy $\cal E$ is varied for a fixed value of $W$,
the distribution of conductance  changes qualitatively and
demonstrates something like the smoothed Anderson transition. The
true phase transition arises in the limit  $W\to 0$, where
$\gamma\to 0$ and $\gamma\to \infty$ in the metallic and
localized phase correspondingly, while the width of the critical
region ($\gamma\sim 1$) tends to zero.

\begin{figure}
\centerline{\includegraphics[width=2.8 in]{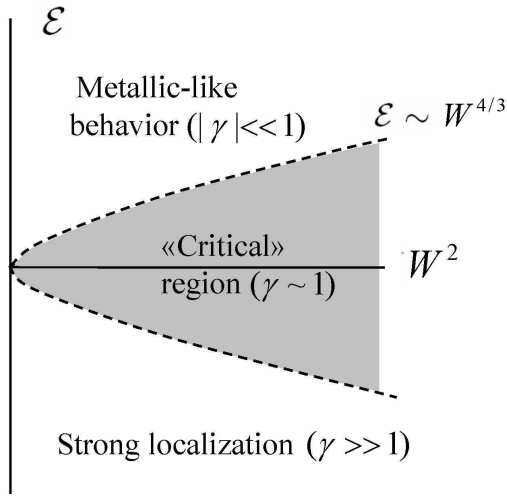}}
\caption{\footnotesize
In the plane $({\cal E},W^2)$ one can distinguish
three characteristic regions: quasi-metallic
($|\gamma|\ll 1$), strongly localized  ($\gamma\gg 1$) and
"critical" ($\gamma\sim 1$). If the energy ${\cal E}$
is changed for a fixed amplitude $W$ of the random potential,
then the conductance distribution undergoes qualitative
changes resembling the smoothed Anderson transition. In the
limit $W\to 0$ the true phase transition arises.
} \label{fig4}
\end{figure}

Let now discuss the dependence of results on the properties
of the ideal leads. We begin with a simple example, in order
to demonstrate the existence of a subject for discussion.
Removing the random potential in the situation of Fig.2,c,
one has a simple quantum mechanical problem on the
tunneling through the potential barrier \cite{24}. Calculating
the amplitudes of transmission ($t$) and reflection ($r$),
one can find the Landauer resistance $\rho=|r|^2/|t|^2$
of the system
$$
\rho= \frac{1}{4} \left(\frac{k}{\kappa}+
\frac{\kappa}{k}\right)^2 {\rm sh}^2{\kappa L}
\,,
\eqno(4)
$$
where $k$ is the Fermi momentum in the ideal leads, and
$\kappa$ is the decrement of decay of the wave
functions under the barrier. The $\rho$ dependence on $L$
is determined by the parameter $\kappa$, i.e. internal properties
of the system, while the proportionality coefficient depends on
$k$, i.e. the properties of the ideal leads.

This result contradicts to a physical intuition, so let us
consider its interpretation. Our physical intuition was
formed on the usual Ohmic regime, when the resistance
$\rho$ is proportional to the system length $L$ or the
number of scatterers $n$; hence, each
scatterer gives the additive contribution to
resistance. However, there exists the localization
regime, where $\rho$ depends on $L$ exponentially, so
$\ln \rho\propto L \propto n$  and each scatterer gives the
multiplicative contribution to resistance. Since the ideal
leads do not provide a dissipation, their contribution to
resistance is related only with interfaces, and has a
multiplicative character in the exponential regime. The latter
changes the common coefficient in (4), which can vary
from unity  (for $k=\kappa$) till infinity.

Generally, the contribution of  interfaces is not additive, nor
multiplicative, and a situation is not trivial (Sec.8). In
addition, the conductance of a finite system is a strongly
fluctuating quantity \cite{25,26}, and one should consider its
distribution function. Dependence of the latter on the boundary
conditions is manifested already in the metallic regime, where it
can be investigated both analytically \cite{27,28}, and
numerically \cite{29}.  \vspace*{2mm}

Dependence of the conductance distribution of 1D systems
on the properties of ideal leads is discussed in
details in Sec.6. Let us declare several simple
statements.
\vspace*{2mm}

1. For large $L$ the distribution $P(\rho)$ is
log-normal,  and contains the limiting value of the parameter
$\gamma$ at $L\to\infty$:
$$
P(\rho)=\frac{1}{\rho \sqrt{4\pi t}}
\exp\left\{-\frac{[\ln \rho-(2\gamma\!+\!1)t]^2}{4t}\right\},
\quad t=\tilde\alpha L \,.
\eqno(5)
$$
It is a solution of Eq.2 for large  $\rho$, when $1+2\rho\approx
2\rho$, $\rho(1+\rho)\approx \rho^2$. The parameters
$\tilde\alpha$ and $\gamma$ can be established (Sec.5) using the
exponents of growth for the second and fourth moments  of the
transfer matrix elements (Sec.4); they are determined by the internal
properties of the system and do not depend on the ideal leads.
The latter affect only the absolute scale of conductance.
\vspace*{2mm}

2. In the deep of the forbidden band (Fig.2,c) the latter
statement remains  valid beyond the log-normal regime:
the ideal leads do not affect the form of the distribution
$P(\rho)$, but change the absolute scale of $\rho$,
$$
P(\rho) \to A P\left(A \rho\right) \,,
\qquad A= 1/\Delta_2^2 \,,
\eqno(6)
$$
where the parameter  $\Delta_2$ is defined below in Eq.9.
  \vspace*{2mm}

3. In the critical region, the situation is more complicated:
the ideal leads change only the absolute scale of conductance,
but the $A$ value in Eq.6 is different  in the log-normal
regime and in the range of not very large $L$. In the latter
case for the "natural" leads (Fig.2,a) one has the following
distribution
$$
P\left(\rho\right) =\frac{1}{\Gamma(\gamma\!+\!1)}
\frac{\rho^\gamma \exp\left\{-\rho/t \right\}}{t^{\gamma+1}} \,,
\eqno(7)
$$
which is a solution of Eq.2 for small $\rho$, when
$1\!+\!2\rho\approx 1$, $\rho(1\!+\!\rho)\approx\rho$. In the
critical region one has $\gamma=-1/2$, while the parameter
$A$ in Eq.6 takes a value $1/\left(\Delta_2\!-\!\Delta_1\right)^2$,
where  $\Delta_1$ and $\Delta_2$ are defined in Eq.9.
  \vspace*{2mm}

4. The situation is even more complicated in the deep of the
allowed band, where the ideal leads affect the whole
distribution function in the range of not very large $L$.
If for the "natural" leads (Fig.2,a)  one has distribution
(7) with $\gamma=0$, then for the foreign leads (Fig.2,b)
its form is essentially modified: in particular, the universal
distribution appears in the formal limit
$L\to 0$:\,\footnote{\,It should be stressed, that the
limit $L\to 0$ is indeed formal, since the results (7)
and (8) are restricted by the condition $L\agt 1/\kappa$.}
$$
P(\rho)=\frac{1}{\pi} \sqrt{\frac{1}{\rho(\rho_c\!-\!\rho)}}
\, \Theta(\rho_c\!-\!\rho) \,,
\eqno(8)
$$
whose evolution with $L$ is demonstrated in Fig.5,a and Fig.5,b;
the latter differ by the values of parameters
$$
\Delta_1=\frac{1}{2}
\left(\frac{k}{\kappa}-\frac{\kappa}{k} \right)\,,\qquad
\Delta_2=\frac{1}{2}
\left(\frac{k}{\kappa}+\frac{\kappa}{k} \right)\,.
\eqno(9)
$$
Parameters (9) are defined for the forbidden band, but
acquire the same form in the allowed band, if one set
formally  $\bar k=\kappa$ for the Fermi momentum $\bar k$
of the system under consideration; they are bounded by
the relation
$$
\Delta_2^2-\Delta_1^2=1 \,,
\eqno(10)
$$
which is of vital significance for conservation of
probability. Below we suggest for convenience that $\Delta_1$
is a free parameter, variating from $-\infty$ till $\infty$,
while the positive parameter  $\Delta_2$ is defined
by Eq.10.

\begin{figure*}
\centerline{\includegraphics[width=6.4 in]{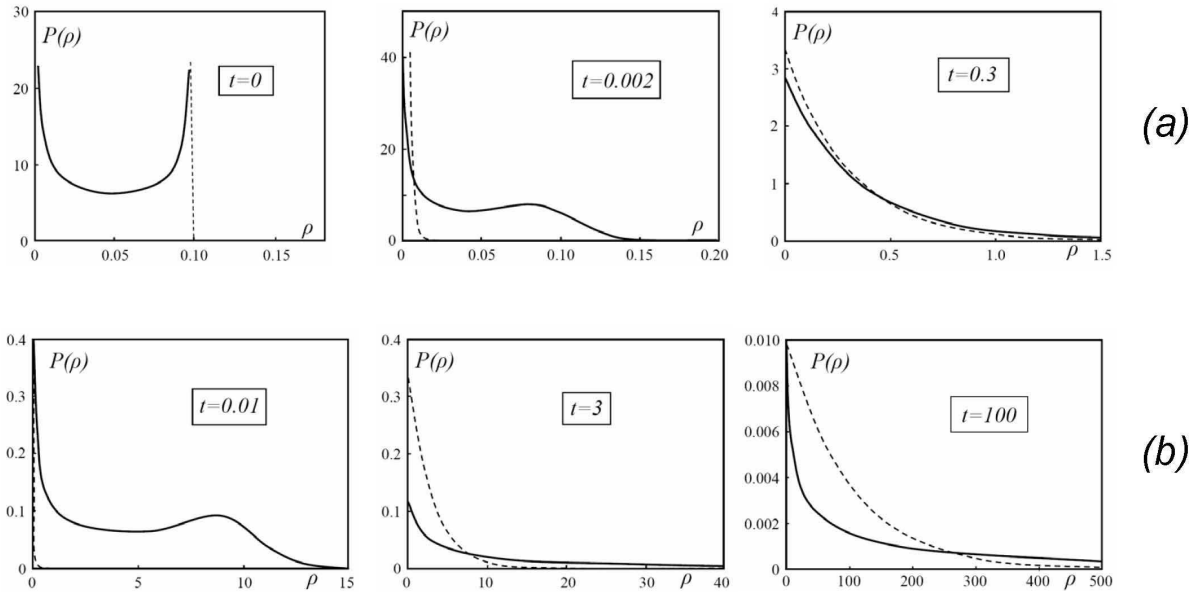}}
\caption{\footnotesize Evolution of the distribution $P(\rho)$ in the
metallic regime for (a) the weak ($\Delta_1^2=0.1$) and (b)
strong ($\Delta_1^2=10$) difference of the Fermi levels
in the given sample and the ideal leads. The dashed line
shows the distribution (7) with $\gamma=0$.
}
\label{fig5}
\end{figure*}

In 1D systems, the mean value of $\rho$ is usually not
representative, but nevertheless is observable (Sec.8).
Its evolution allows a complete description for arbitrary
$L$ and gives a clear demonstration for the influence of
the ideal leads. As function of $L$, both
mean of $\rho$, and its higher moments can exhibit
incommensurate oscillations, which provide the aperiodic
character of oscillations for the resistance $\rho$ in the
specific sample (Fig.8).

\begin{center}
{\bf 2. Different types of transfer matrices} \end{center}

The use of transfer matrices is the natural approach
to investigation of 1D systems. The most convenient variant is
the transfer matrix $T$ in the wave representation, which relates
the amplitudes of waves on the left ($Ae^{ikx}+Be^{-ikx}$) and on
the right ($Ce^{ikx}+De^{-ikx}$)  of a scatterer:
$$
\left ( \begin{array}{cc} A \\ B \end{array} \right)\,
=  T \left ( \begin{array}{cc} C \\ D \end{array}
\right) \,.
\eqno(11)
$$
It is determined by the transmission ($t$)
and reflection ($r$) amplitudes   and in the presence of the
time-reversal invariance allows the following parametrization
\cite{1}
$$
 T= \left ( \begin{array}{cc} \!\!1/t & - r/t \!\!\\
\!\!- r^*/t^* & 1/t^* \!\!\end{array} \right)\,
= \left ( \begin{array}{cc}\! \sqrt{\rho\!+\!1}\, e^{i\varphi}\!\! &
\!\!\sqrt{\rho}\! \,e^{i\theta}
\\ \!\sqrt{\rho}\, e^{-i\theta}\!\! & \!\!\sqrt{\rho\!+\!1}\,
e^{-i\varphi} \!
\end{array} \right)\,,
\eqno(12)
$$
where $\rho=|r/t|^2$ is the Landauer resistance \cite{2}. For the
successive arrangement of scatterers their transfer matrices
are multiplicated. For a weak scatterer, the matrix $T$ is
close to the unit one, which allows to derive the differential
evolution equations for its elements and the Landauer
resistance $\rho$.

For the energy in the forbidden band wave functions on the
left ($Ae^{\kappa x}+Be^{-\kappa x}$) and on the right
($Ce^{\kappa x}+De^{-\kappa x}$) of a scatterer are given by a
superposition of the increasing and decreasing exponents
and can be related by the pseudo-transfer matrix
$$
\left ( \begin{array}{cc} A \\ B \end{array} \right)\,
= {\cal T} \left ( \begin{array}{cc} C \\ D \end{array}
\right) \,= \left ( \begin{array}{cc} t_{11} & t_{12} \\
t_{21} & t_{22} \end{array} \right)\,
\left ( \begin{array}{cc} C \\ D \end{array}
\right) \,,
\eqno(13)
$$
for which parametrization (12) is not valid and its relation
with $\rho$ is not evident. The elements of the matrix ${\cal T}$
are real and its determinant is equal to unity, as in the case of
(12).

At last, one can introduce the transfer matrix in the
coordinate representation, as can be illustrated for
the 1D Anderson model, describing by the discrete
Schroedinger equation
$$
\Psi_{n+1}+\Psi_{n-1}+V_n \Psi_n = E \Psi_n \,,
\eqno(14)
$$
where $E$ is the energy counted from the band center,
and the hopping integral is set to be unity. Rewriting (14)
in the form
$$
\left ( \begin{array}{cc} \!\Psi_{n+1} \!
\\ \!\Psi_n \!\end{array} \right)\,
= \left ( \begin{array}{cc} E-V_n & -1 \\
1 & 0 \end{array} \right)\,
\left ( \begin{array}{cc}\! \Psi_n \! \\ \!\Psi_{n-1}\!
\end{array}
\right) \,,
\eqno(15)
$$
and making $n$ iterations, one can obtain
$$
\left ( \begin{array}{cc}\! \Psi_{n+1} \!
\\ \! \Psi_n\! \end{array} \right)\,=\tau\,
\left ( \begin{array}{cc}\! \Psi_1\! \\ \!\Psi_0 \!\end{array}
\right) \,
= \left ( \begin{array}{cc}\!\! \tau_{11}\! & \!\tau_{12}\!\!\\
\!\!\tau_{21}\! & \!\tau_{22}\!\! \end{array} \right)\,
\left ( \begin{array}{cc} \!\Psi_1\! \\ \!\Psi_0\! \end{array}
\right) \,,
\eqno(16)
$$
where the matrix ${\tau}$ is a product of $n$ matrices of
type (15).

Three matrices $T$, ${\cal T}$, $ \tau$ are determined
by the internal properties of the system, and have their merits
and drawbacks (Fig.6). These matrices do not allow to obtain the
differential evolution equation for $\rho$, applicable
for all energies. Indeed, the matrix $T$ possesses
the necessary properties, but immediately applicable only in
the allowed band; in the forbidden band its role is played by
the pseudo-transfer matrix ${\cal T}$, which has no direct
relation with $\rho$. The matrix $\tau$ is applicable for all
energies, but has no direct relation with $\rho$ and is not close
to the unit one for a weak scatterer. The matrices ${\cal T}$ and
$\tau$ consist of the real elements, which has some technical
advantages\,\footnote{\,In particular, the analysis of
fourth moments for matrices with the complex
elements looks rather hopeless, since it demands diagonalization
of the matrix of large size.}.

\begin{figure*}
\centerline{\includegraphics[width=5.0 in]{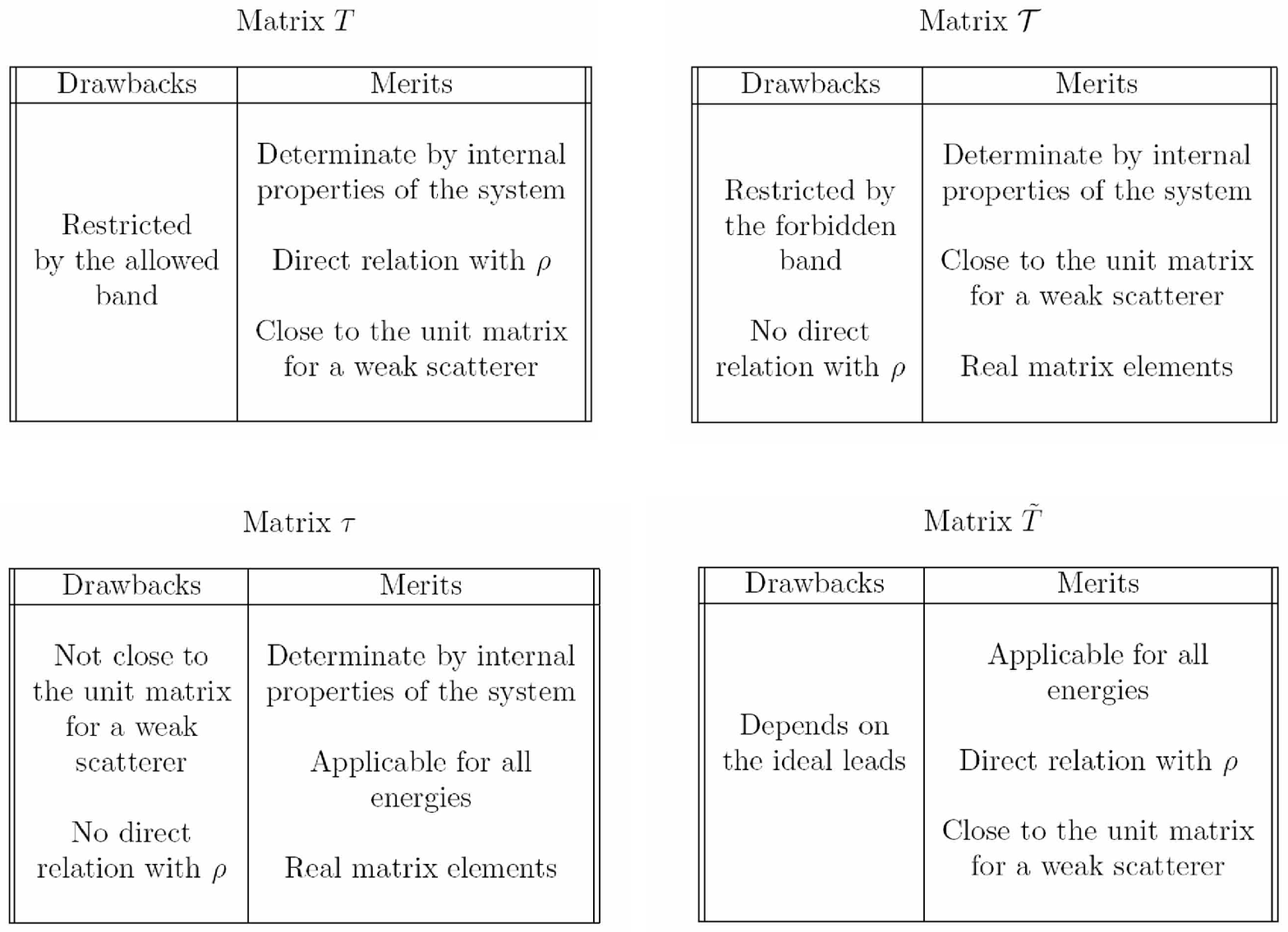}}
\caption{\footnotesize Merits and drawbacks
of different transfer matrices.
} \label{fig6}
\end{figure*}

To overcome these difficulties, let introduce the "external"
matrix $\tilde T$, which takes into account the presence
of the foreign leads (Fig.2,b) and is related with the "internal"
matrix $T$ of the system with the help of the "edge" matrices
$T_a$ and $T_b$:
$$
\tilde T= T_a T T_b=
 \left ( \begin{array}{cc} \! \! a \!&\! a_1 \!\! \\
 a_1\! &\! \! \! a \! \!\end{array} \right)
 \left ( \begin{array}{cc}\! T_{11} & T_{12}\! \\
\!T_{21} & T_{22}\! \end{array} \right)
 \left ( \begin{array}{cc} \! \! b\! &\! b_1\!\!  \\
b_1\! &\! \! \! b\!\!  \end{array} \right),
$$
$$
a=\frac{1}{2}\left(1+\frac{\bar k}{k} \right)\,,\qquad
a_1=\frac{1}{2}\left(1-\frac{\bar k}{k} \right)\,,\qquad
\eqno(17)
$$
$$
b=\frac{1}{2}\left(1+\frac{k}{\bar k} \right)\,, \qquad
b_1=\frac{1}{2}\left(1-\frac{k}{\bar k} \right)\,.
$$
The edge matrices have a simple sense: for example, $T_a$
relates the amplitudes of waves on the left
of the interface ($Ae^{ikx}+Be^{-ikx}$) and on the right
of it ($A_1 e^{i \bar k x}+B_1 e^{-i\bar k x}$). One can see
that $T_a \, T_b=1$, i.e. the edge matrices are are mutually
inverse.

For the energy in the forbidden band (Fig.2,c) the explicit
introduction of ideal leads is inevitable, and is also
realized with the help of the edge matrices, transforming
the pseudo-transfer matrix ${\cal T}$ to the true transfer matrix
$\tilde T$
$$
\tilde T=  T_a \,{\cal T}\,  T_b=
 \left ( \begin{array}{cc} \!\! a &\! a^* \!\! \\
a^* &\! \!\! a \!\!\end{array} \right)
 \left ( \begin{array}{cc} \! t_{11} & t_{12}\! \\
\!t_{21} & t_{22}\! \end{array} \right)
 \left ( \begin{array}{cc} \!\!\! b  &\! b^* \!\!\\
\!b^* &\! \!\! b\!\! \end{array} \right),
$$
$$
a=\frac{1}{2}\left(1+\frac{\kappa}{ik} \right)\,,\qquad
b=\frac{1}{2}\left(1+\frac{ik}{\kappa} \right)\,.
\eqno(18)
$$
Relations (18) can be obtained from (17) by the change
$\bar k\to -i\kappa$.

The use of the edge matrices allows also to relate
$\tilde T$ and $\tau$
$$
\tilde T= \frac{1}{2i \sin{ka_0}}
 \left ( \begin{array}{cc} 1 & -{\rm e}^{-ika_0} \\
-1 & {\rm e}^{ika_0} \end{array} \right)\, \cdot
\qquad\qquad
$$
$$ \qquad\qquad \cdot
 \left ( \begin{array}{cc} \tau_{11} & \tau_{12} \\
\tau_{21} & \tau_{22} \end{array} \right)\,
 \left ( \begin{array}{cc} {\rm e}^{ika_0} & {\rm e}^{-ika_0} \\
1 & 1 \end{array} \right)\,,
\eqno(19)
$$
where $a_0$ is the lattice constant for the model (14). One can
see that the explicit introduction of the ideal leads
corresponds to the
linear transformation of matrix elements. A linear
transformation does not change the exponents of growth for the
second and fourth moments, which are determined by the internal
properties of the system and do not depend on the ideal leads
(Sec.4). As should be clear from Fig.6, the matrix $\tilde T$
possesses all necessary properties, but depends on the ideal
leads; the latter can be consider as a drawback, but in
fact it reflects the objective reality.

\begin{center}
{\bf 3. Succession of point scatterers}
\end{center}

The coefficients  $\tilde\alpha$, $\gamma_1$,  $\gamma_2$ of
equation (3) are not necessary constant and can variate in the
course of evolution. To gain an insight into the character of this
evolution, let consider a succession of point scatterers.
For definiteness, let have in mind the Anderson model (14)
near the band edge, where it corresponds to discretization of
the usual Schroedinger equation.

One can easily verify that the point scatterer with the
potential $V\delta_{n0}$ is described by the
transfer matrix
$$
\left ( \begin{array}{cc} 1-i\epsilon & -i\epsilon \\
i\epsilon & 1+i\epsilon \end{array} \right)\,,\qquad
\mbox{\rm where}\quad \epsilon =\frac{V}{2ka_0}\,.
\eqno(20)
$$
If two scatterers with amplitudes $V_0$ and $V_1$ are arranged at
the distance $L$, then the transfer matrix arises
$$
\left ( \begin{array}{cc} 1-i\epsilon_0 & -i\epsilon_0 \\
i\epsilon_0 & 1+i\epsilon_0 \end{array} \right)\,
\left ( \begin{array}{cc} {\rm e}^{-ikL} & 0 \\
0 & {\rm e}^{ikL} \end{array} \right)\, \cdot \qquad\qquad
$$
$$
\qquad\qquad\qquad\qquad\,\,\cdot
\left ( \begin{array}{cc} 1-i\epsilon_1 & -i\epsilon_1 \\
i\epsilon_1 & 1+i\epsilon_1 \end{array} \right)\,.
\eqno(21)
$$
At last, if the scatterers with amplitudes
$V_0$,\,$V_1$,\,$V_2$,\,$\ldots$,\,$V_n$ are arranged at the
points $0$, $L_1$, $L_1+L_2$,\,$\ldots$, then the corresponding
transfer matrix has a form
$$
 T^{(n)}= T_{\epsilon_0}
\, T_{\delta_1} \, T_{\epsilon_1}\,
T_{\delta_2}\,T_{\epsilon_2}\,
\ldots\,
 T_{\delta_n}\, T_{\epsilon_n}\,,
\eqno(22)
$$
where
$$
 T_{\epsilon_s}=
\left ( \begin{array}{cc} 1-i\epsilon_s & -i\epsilon_s \\
i\epsilon_s & 1+i\epsilon_s \end{array} \right)\,,\quad
\epsilon_s =\frac{V_s}{2ka_0}\,,
$$
$$
T_{\delta_s}= \left ( \begin{array}{cc}
{\rm e}^{-i\delta_s} & 0 \\ 0 & {\rm e}^{i\delta_s} \end{array}
\right)\,,\quad
\quad \delta_s =kL_s\,.
\eqno(23)
$$
If the number of scatterers is increased by  unity, one has the
recurrent relation
$$
T^{(n)} =T^{(n-1)}
 \left ( \begin{array}{cc} u_n\,\, & v_n \\
v^*_n & u^*_n\,\, \end{array} \right)\,,\qquad
\eqno(24)
$$
$$
u_n=\left(1-i\epsilon_n\right)\,{\rm e}^{-i\delta_n}\,,
\qquad
v_n=-i\epsilon_n\,{\rm e}^{-i\delta_n}\,.
$$
For convenience we set $\epsilon_0=0$, in order to use
the unit matrix as the initial condition for $T^{(0)}$. If all
$\epsilon_n$ are small, then for not very large $n$ one
can retain two first orders in $v_n$; producing
direct multiplication of matrices, one has
$$
T^{(n)}_{11}= u_1 u_2 \ldots u_n +
\sum\limits_{i=1}^{n-1}
\sum\limits_{j=i+1}^{n} u_1 u_2 \ldots u_{i-1} v_i \,\cdot
$$
$$
\qquad\qquad\qquad\qquad\cdot \,
u^*_{i+1} \ldots u^*_{j-1} v^*_{j} u_{j+1} \ldots u_n \,,
$$
$$
T^{(n)}_{12}= \sum\limits_{i=1}^{n}
u_1 u_2 \ldots u_{i-1} v_i
u^*_{i+1} \ldots u^*_{n}  \,
\eqno(25)
$$
and $T_{21}=T^*_{12}$, $T_{22}=T^*_{11}$.

In the Anderson model (14), a scatterer is present at each
site of the lattice, so all $\delta_s$ are equal,
$\delta_s=ka_0$. Expressions (25) can be reduced to the
form
$$
T^{(n)}_{11}= \sqrt{1+S_1^2+S_2^2}\, {\rm e}^{-iS-in\delta}\,,
\quad
 T^{(n)}_{12}= S_1-iS_2 \,,
\eqno(26)
$$
where in the main order in $\epsilon$
$$
S= \sum\limits_{s=1}^{n} \epsilon_s\,,\quad
S_1= \sum\limits_{s=1}^{n} \epsilon_s \sin{(n\!-\!2s)\delta}\,,
\quad
$$
$$
S_2= \sum\limits_{s=1}^{n} \epsilon_s \cos{(n\!-\!2s)\delta}\,.
\eqno(27)
$$
Accepting as usually for the Anderson model
$$
\langle V_n \rangle= 0\,,\qquad \langle V^2_n
\rangle =W^2\,,
\eqno(28)
$$
one has the zero means for all sums in (27), while for
the second moments
$$
\langle S^2 \rangle=2\langle S_1^2 \rangle=
2\langle S_2^2 \rangle= \epsilon^2 n \,,\qquad
$$
$$
\langle S S_1 \rangle\sim \langle S S_2 \rangle \sim
\langle S_1 S_2 \rangle\sim \epsilon^2\,,
\eqno(29)
$$
where  $\epsilon^2=W^2/4k^2 a_0^2$. Above expressions
are valid under condition
$$
1/\delta \ll n \ll 1/\epsilon^2 \,,
\eqno(30)
$$
where the left inequality provides the large number of
oscillations of sine and cosine in sums (27), while the right
inequality is necessary for neglection of higher orders in
$\epsilon$. For large  $n$, all sums (27) are normally
distributed and practically uncorrelated, so their mutual
distribution function has a form
$$
P\left(S, S_1, S_2 \right) \sim
\exp\left\{-\frac{S^2}{2n\epsilon^2}
-\frac{S_1^2}{n\epsilon^2} -\frac{S_2^2}{n\epsilon^2}
\right\}\,.
\eqno(31)
$$
Using the relations $\rho=S_1^2+S_2^2$, $\varphi=-S-n\delta$,
one has the mutual distribution for the elements of the
transfer matrix  (12)
$$
P\left(\rho, \varphi, \theta \right) \sim
\exp\left\{-\frac{(\varphi+n\delta)^2}{2n\epsilon^2}
-\frac{\rho}{n\epsilon^2} \right\}\,.
\eqno(32)
$$
It should be clear,
that for $\delta\gg \epsilon^2$, i.e. in the deep
of the allowed band, the phase $\theta$ becomes
completely random on the scale $n\sim 1/\delta$. Stochastization
of the phase $\varphi$ occurs on the scale $n\sim 1/\epsilon^2$,
but its uniform distribution arises
already on the scale $1/\delta$
due to regular variations. In fact, stochastization of
$\theta$ is sufficient for applicability of the random
phase approximation and disappearance of $\gamma$,
since the evolution equation for $P(\rho)$
contains only the combination
$\psi=\theta-\varphi$ (see Sec.7). Solution of Eq.2 for small
$\rho$ has a form (7), which agrees with (32) for $\gamma=0$
and $t=\epsilon^2n$; the latter relations follow
from Eq.93 of Sec.7.  For large  $n$ the
exponential growth of the elements $T_{ij}$ is
determined by the product
$|u_1||u_2|\ldots |u_n|\equiv {\rm e}^{w_n}$, which
is contained
in all terms of (25), as well as in the higher order terms.
The quantity $w_n$ has the mean value $\epsilon^2 n/2$,
which (in view $\rho=|T_{12}|^2$) agrees with the result
$a=\epsilon^2$ for the parameter $a$ in the log-normal
distribution  (Sec.\,5). Neglected terms do not give
contributions $O(\epsilon^2)$ in the mean of $w_n$,
but change its variance.

Above considerations are valid in the deep of the
allowed band for the "natural" ideal leads (Fig.2,a). The
forbidden band is described by the transfer matrix
$\tilde T^{(n)}= T_a \, {\cal T}^{(n)}\, T_b$,  where
the pseudo-transfer matrix $ {\cal T}^{(n)}$ is determined by
the above relations with the change  $k\to -i\kappa$;
in particular
$$
{\cal T}^{(n)} = {\cal T}^{(n-1)}
 \left ( \begin{array}{cc} u_n\,\, & v_n \\
v^*_n & u^*_n\,\, \end{array} \right)\,,\qquad
\eqno(33)
$$
$$
u_n=\left(1+\epsilon_n\right)\,{\rm e}^{-\delta_n}\,,
\qquad
v_n=\epsilon_n\,{\rm e}^{-\delta_n}\,,
$$
$$
\epsilon_n =V_n/2\kappa a_0\,,
\quad \delta_n =\kappa L_n\,,
$$
where the star indicates not the complex conjugation, but a
change of signs for $\epsilon_n$ and $\delta_n$. With this
modification, relations (25) remain formally valid;
extracting the factor  $ u^*_1 u^*_2
\ldots u^*_n\equiv \exp{(w_n)}$ from all sums, one obtains
$$
{\cal T}^{(n)}=
\left ( \begin{array}{cc}
{\rm e}^{-\tilde w_n} - {\rm e}^{\tilde w_n} S_1
S_2\,\,
& {\rm e}^{\tilde w_n} S_1 \\
-{\rm e}^{\tilde w_n}  S_2
 & {\rm e}^{\tilde w_n} \end{array} \right)\,,\qquad
\eqno(34)
$$
$$
S_1=\sum\limits_{s=1}^{n} \epsilon_{s}
{\rm e}^{-2s\delta} \,, \qquad
S_2=\sum\limits_{s=0}^{n-1} \epsilon_{n-s}
{\rm e}^{-2s\delta}  \,, \qquad
$$
where $\tilde w_n$ differs from $w_n$ by the contribution
$O(\epsilon^2)$ with a zero mean, while  $S_1$ and
$S_2$ are given in the main order in $\epsilon$. Substitution
to (18) gives for the phase variables
$$
{\rm tg}\varphi =-\Delta_1-\Delta_2(S_1+S_2)\,,\qquad
$$
$$
{\rm ctg}\theta =(S_1-S_2)/\Delta_2\,,
\eqno(35)
$$
where $\Delta_1$ and $\Delta_2$ are defined in (9). Having in
mind that under condition (30)
$$
\langle S_1^2 \rangle=\langle S_2^2 \rangle=
\epsilon^2/4\delta\,,\qquad \langle S_1 S_2 \rangle\approx 0
\,,
\eqno(36)
$$
we see that fluctuations of $\varphi$ and $\theta$ are
restricted and do not increase with $n$. The case
$k=\kappa$ is special, since  $\Delta_1=0$ and
$\Delta_2=1$, so
$$
\varphi =-( S_1+S_2)\,,\qquad
\theta = \pi/2+(S_2-S_1)\,,
\eqno(37)
$$
and the variable $\psi=\theta-\varphi$ does not have an
essential evolution, being localized near $\pi/2$ for all
$n\alt 1/\epsilon^2$. As shown in Sec.7, this property
remains valid for $n\agt 1/\epsilon^2$.  For large
$n$ the exponent ${\rm e}^{\tilde w_n}$ provides the
growth of elements $t_{ij}$ and correspondingly
$\rho\sim {\rm e}^{ 2\tilde w_n}$; the quantity $2\tilde
w_n$ has the mean  $(2\delta-\epsilon^2) n$
and the variance $4\epsilon^2 n$, in agreement with Eq.56
of Sec.5.

Let come to the critical region, determined by the condition
$\delta\ll \epsilon^2$. Then for sufficiently small $n$
one can set  $\delta_s=0$  in Eq.23 and obtain
$$
T^{(n)}=
\left ( \begin{array}{cc} 1-i S & -i S \\
i S & 1+i S \end{array} \right),\quad
S= \sum\limits_{s=1}^{n} \epsilon_s\,,
\eqno(38)
$$
where $S$ has the Gaussian distribution analogously
to (31). The mutual distribution for  $\rho$, $\varphi$, $\theta$
is given by the relation
$$
P(\rho, \varphi, \theta)= \int dS \frac{1}{\sqrt{2\pi n
\epsilon^2}} \exp\left(-\frac{S^2}{2 n \epsilon^2}\right)
\cdot
$$
$$ \quad\qquad\cdot  \,
\delta\left(\rho-S^2\right)\,
\delta\left(\varphi+{\rm arctg} S\right)\,
\delta\left(\theta+\frac{\pi}{2}\right)\,
\eqno(39)
$$
and integration over $\varphi$ and $\theta$ gives the result
$$
P(\rho)=  \frac{1}{\sqrt{\rho}}\frac{1}{\sqrt{2\pi n
\epsilon^2}} \exp\left(-\frac{\rho}{2 n \epsilon^2}\right)
\,,
\eqno(40)
$$
which agrees with (7) for $\gamma=-1/2$, $t=2n\epsilon^2$;
the latter follows from Eq.93 of Sec.7, if $\psi$ is localized
near $\pm \pi/2$. Such localization is indeed valid for
small $\rho$, when $\varphi\sim \sqrt{\rho}\ll 1$,
$\theta=-\pi/2$, and (40) is a solution of Eq.2.
However, the result (40) remains valid in the more wide
interval $n\alt (\epsilon^2\delta^2)^{-1/3}$ (Sec.8), when
$\rho\sim n\epsilon^2$  can be large; in this case
$\varphi$ is localized near $\pm \pi/2$, while $\sin{\psi}$
becomes small in agreement with the results for the
log-normal regime (Secs.5,\,7).

\begin{center}
{\bf 4. Evolution of moments}
\end{center}

According to (33), the elements of the pseudo-transfer matrix in
the forbidden band obey the evolution equations
$$
t_{11}^{(n)}= u_n t_{11}^{(n-1)} + v_n^* t_{12}^{(n-1)}\,,\quad
t_{12}^{(n)}= v_n t_{11}^{(n-1)} + u_n^* t_{12}^{(n-1)}\,,
\eqno(41)
$$
\onecolumn

\noindent
and the analogous equations for $t_{21}^{(n)}$ and $t_{22}^{(n)}$;
it is essential that $t_{ij}^{(n-1)}$ do not contain $\epsilon_{n}$
and are statistically independent
of $u_n$, $v_n$. Introducing the notations for the second
moments
$$
z_{1}^{(n)}=\left\langle
\left[t_{11}^{(n)}\right]^2\right\rangle\,, \qquad
z_{2}^{(n)}=\left\langle
t_{11}^{(n)}t_{12}^{(n)}\right\rangle\,, \qquad
z_{3}^{(n)}=\left\langle
\left[t_{12}^{(n)}\right]^2\right\rangle\,,
 \eqno(42)
$$
one can obtain the system of the linear difference equations
with constant coefficients
$$
\left ( \begin{array}{ccc} z_{1}^{(n)} \\ z_{2}^{(n)}
\\ z_{3}^{(n)}\end{array} \right)\,
= \left ( \begin{array}{ccc} 1-2\delta +\epsilon^2 &
-2\epsilon^2 & \epsilon^2
\\ \epsilon^2 & 1-2\epsilon^2 & \epsilon^2 \\
\epsilon^2 & -2\epsilon^2 & 1+2\delta+ \epsilon^2
 \end{array} \right)\,
\left ( \begin{array}{ccc} z_{1}^{(n-1)} \\ z_{2}^{(n-1)}
\\z_{3}^{(n-1)} \end{array} \right) \,,
\eqno(43)
$$
whose solution is sought in the exponential form,
$z_{i}^{(n)} \sim \lambda^n $ \cite{30}; it is easy to
see that $\lambda$ is an eigenvalue of the matrix (43).
Setting $\lambda=1+x$, one has the equation for $x$
$$
x\left(x^2-4\delta^2\right)= 8\epsilon^2 \delta^2 \,.
\eqno(44)
$$
We have in mind the limiting transition
$$
\delta\to 0\,,\qquad \epsilon\to 0\,, \qquad
\delta/\epsilon^2=const \,,
\eqno(45)
$$
which is actual near the band edge of the ideal crystal,
where one can neglect the effects of commensurability
\cite{32,33} complicating the analysis; correspondingly,
we retain the terms of the first order in
$\delta$ and the second order in $\epsilon$ in the matrix
(43).

Analogously we set for the fourth moments
$$
z_{1}^{(n)}=\left\langle
\left[t_{11}^{(n)}\right]^4\right\rangle\,, \qquad
z_{2}^{(n)}=\left\langle
\left[t_{11}^{(n)}\right]^3 t_{12}^{(n)}\right\rangle\,, \qquad
z_{3}^{(n)}=\left\langle \left[t_{11}^{(n)}\right]^2
\left[t_{12}^{(n)}\right]^2\right\rangle\,,
$$
$$
z_{4}^{(n)}=\left\langle t_{11}^{(n)}
\left[t_{12}^{(n)}\right]^3\right\rangle\,, \qquad
z_{5}^{(n)}=\left\langle
\left[t_{12}^{(n)}\right]^4\right\rangle\,,
 \eqno(46)
$$
and obtain the system of difference equations
$$
\left ( \begin{array}{ccccc} z_{1}^{(n)} \\ z_{2}^{(n)}
\\ z_{3}^{(n)} \\ z_{4}^{(n)} \\ z_{5}^{(n)}\end{array} \right)\,
= \left ( \begin{array}{ccccc} 1-4\delta +6\epsilon^2 &
-12\epsilon^2 & 6\epsilon^2 &  0& 0\\
3\epsilon^2  & 1-2\delta-3\epsilon^2 & -3\epsilon^2 &
3\epsilon^2 & 0 \\
\epsilon^2 & 2\epsilon^2 & 1-6\epsilon^2 & 2\epsilon^2
& \epsilon^2 \\
0 & 3\epsilon^2 & -3\epsilon^2 & 1+2\delta -3\epsilon^2
& 3\epsilon^2 \\
0 & 0 & 6\epsilon^2 &  -12\epsilon^2 & 1+4\delta+6\epsilon^2
 \end{array} \right)\,
\left ( \begin{array}{ccc} z_{1}^{(n-1)} \\ z_{2}^{(n-1)}
\\z_{3}^{(n-1)} \\z_{4}^{(n-1)} \\z_{5}^{(n-1)}\end{array}
\right) \,,
\eqno(47)
$$
Accepting $z_{i}^{(n)} \sim \lambda^n$ and setting
$\lambda=1+x$, we have the equation for $x$
$$
x\left(x^2-4\delta^2\right)\left(x^2-16\delta^2\right)=
24\epsilon^2 \left(7\delta^2 x^2-16 \delta^4\right ) \,.
\eqno(48)
$$

Curiously, equations (44) and (48) can be obtained, if one
composes the product of diagonal elements of matrices (43), (47)
and retains the terms of the second order in $\epsilon$. Indeed,
non-diagonal elements give contributions $O(\epsilon^4)$ and
higher, whose
concellation can be foreseen beforehand. Since
$\epsilon^2=W^2/4\delta^2$ (see Sec.3), the combinations
$\epsilon^2$, $\epsilon^4$, ..., $\epsilon^{10}$ and
$\delta^2\epsilon^4$, $\delta^2\epsilon^6$ contain
singularities at $\delta\to 0$, whose absence is evident from
the evolution equations for the coordinate transfer matrix (see
Appendix 1); only combinations $\delta^2\epsilon^2$ and
$\delta^4\epsilon^2$ are allowable,  which enters in
Eqs.44,\,48.\,\footnote{\,We have listed all possible
combinations. Indeed, a change of the $\delta$ sign in
(43), (47) leads to the analogous matrices, which can be
transformed to  the initial form, if the components of columns are
renumerated in the inverse order; so the odd powers of
$\delta$ do not appear. Since we consider the limit
$\delta\sim\epsilon^2\to 0$ (see Eq.45), then only combinations
$\delta^{2n}\epsilon^{2m}$ with $4n+2m\le 6$ for (43) and
$4n+2m\le 10$ for (47) are possible; among them only
combinations with  $n\ge m$ do not have singularities for
$\delta\to 0$.} In the deep of the allowed or forbidden band,
the possibility of restriction by diagonal elements allows to
establish the exponents of growth for higher moments and verify
their accordance to the log-normal distribution.
\twocolumn

\noindent

For the Anderson model (14) one can set $\delta^2=-\cal E$,
$4\epsilon^2\delta^2=W^2$, where $\cal E$ is the energy counted
from the lower band edge; correspondingly, equations (44), (48)
can be rewritten in the form
$$
x\left(x^2 +4{\cal E}\right)= 2W^2
\eqno(49)
$$
$$
x\left(x^2+{4\cal E} \right)\left(x^2 +16{\cal E}\right)=
42W^2 x^2+96 W^2 {\cal E}  \,.
$$
Equations (49) were derived for ${\cal E}<0$, but can be
analytically continued to  arbitrary $\cal E$ due to
regularity in $\cal E$. These equations can be derived also
using the coordinate transfer matrix (see Appendix 1),
which is applicable for arbitrary  $\cal E$ and does not require
the analytical continuation. One of the roots for each of
equations (49) remains positive for all physical values
%
%
of parameters; it has the maximal real part between all
roots of the equation and determines the exponent of growth
for the second  ($x_2$) or fourth ($x_4$) moments.
Behavior of $x_2$ and $x_4$ against ${\cal E}/W^{4/3}$ is
shown in Fig.7.

\begin{figure}
\centerline{\includegraphics[width=2.8 in]{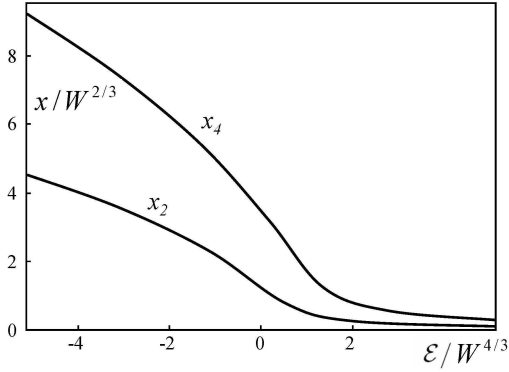}}
\caption{
The positive root of the first ($x_2$) and
second ($x_4$) equation (49) against the energy
${\cal E}$.
} \label{fig7}
\end{figure}

It is not difficult to obtain the asymptotic results for $x_2$
and $x_4$. In the deep of the forbidden band
($\delta\gg\epsilon^2$) we have from (44), (48)
$$
x_2=2\delta+\epsilon^2\,,\quad x_4=4\delta+6\epsilon^2\,.
\eqno(50a)
$$
To come into the allowed band, one makes the replacements
$\delta\to i\delta$, $\epsilon\to -i\epsilon$ in Eqs.44,\,48,
and in the deep of the band  ($\delta\gg\epsilon^2$) finds
the results
$$
x_2=2\epsilon^2\,, \quad
x_4=6\epsilon^2\,.
\eqno(50b)
$$
In the critical region ($\delta\ll\epsilon^2$) one has
$$
x_2=2\left(\epsilon^2\delta^2\right)^{1/3} \,,\quad
x_4=2\left(21\epsilon^2\delta^2\right)^{1/3}\,.
\eqno(50c)
$$
In terms of ${\cal E}$ and  $W$ we can write
$$
x_2=2|{\cal E}|^{1/2}+W^2/4|{\cal E}|\,, \quad
 \qquad\qquad\qquad\qquad
$$
$$
x_4=4|{\cal E}|^{1/2}+3W^2/2|{\cal E}|
\mbox{\quad for }  -{\cal E}\gg W^{4/3} \,,
$$
$$
x_2=\left(2W^2\right)^{1/3},\,\,
x_4=\left(42W^2\right)^{1/3}
\mbox{\,\, for \,}  |{\cal E}|\ll W^{4/3}  \,,
\eqno(51)
$$
$$
x_2=W^2/2{\cal E} -W^6/32{\cal E}^4\,, \quad
\qquad\qquad\qquad\qquad
$$
$$
x_4=3W^2/2{\cal E}+27W^6/64{\cal E}^4\,
\mbox{\quad for }  {\cal E}\gg W^{4/3} \,,
$$
where the latter result is given with higher accuracy.

\vspace{3mm}

\begin{center}
{\bf 5. Parameters of the log-normal distribution}
\end{center}

For large $L=na_0$ the distribution $P(\rho)$ is
log-normal
$$
P(\rho)=\frac{1}{\rho \sqrt{2\pi bn}}
\exp\left\{-\frac{(\ln \rho-an)^2}{2bn}\right\}\,,
\eqno(52)
$$
as was established in many papers  (\cite{1},
\cite{13}--\cite{18}, \cite{21}--\cite{24}) for partial cases,
and follows most generally from the modified
Dorokhov--Mello--Pereyra--Kumar equation \cite{31}; in the
presence of foreign leads it is derived in Sec.7.
The moments of the distribution (52) can be easily found,
$\left\langle \rho^m\right\rangle=
{\rm e}^{amn+bm^2 n/2}$.
Since $\rho$ is
determined by the expression, quadratic in the transfer matrix
elements, the parameters $a$ and $b$ can be established by
comparison with the growth of the second and fourth moments
$$
\left\langle \rho\right\rangle= {\rm e}^{an+\frac{1}{2}bn}=
C_1{\rm e}^{x_2 n}\,,
$$
$$
\left\langle \rho^2\right\rangle= {\rm e}^{2an+2bn}=
C_2{\rm e}^{x_4 n}\,.
\eqno(53)
$$
It is easy to see that
$$
a=2x_2-\frac{1}{2}x_4\,,\qquad
b=x_4-2x_2\,,
\eqno(54)
$$
while for the parameter $\gamma$ in (5)
$$
\gamma= \frac{a}{b} - \frac{1}{2} =\frac{3x_2-x_4}{x_4-2x_2}
\,.
\eqno(55)
$$
Substitution of the asymptotic expressions (50) gives
$$
a=2\delta-\epsilon^2\,,\quad b=4\epsilon^2\,
\quad\mbox{(forbidden band)}
$$
$$
a=\epsilon^2\,, \quad b=2\epsilon^2\,
\qquad\mbox{(allowed band)}
\eqno(56)
$$
$$
a=1.24\left(\epsilon^2\delta^2\right)^{1/3} ,\quad
b=1.52\left(\epsilon^2\delta^2\right)^{1/3}
\,\,\mbox{(critical region)}
$$
The use of asymptotic forms (51) gives for the parameter
$\gamma$ in  (5)
$$
\gamma=\left \{ \begin{array}{ccc}
2|{\cal E}|^{3/2}/W^2\,,&  -{\cal E}\gg W^{4/3} \\
0.316   \,,&
|{\cal E}|\ll W^{4/3} \\
-33 W^{4}/32{\cal E}^{3}\,, & {\cal E}\gg W^{4/3}
\end{array} \right.\,,
\eqno(57)
$$
while the parameter $\tilde\alpha$ is related with $b$
according to $\tilde\alpha=b/2a_0$.

As shown in Sec.2, introduction of the foreign leads
results in the linear transformation for the elements of the
transfer matrix, which does not change the exponents $x_2$ and
$x_4$  for the second and fourth moments. By this reason, the
latter do not depend on the properties of the ideal leads, as
well as parameters $a$, $b$, $\gamma$. The ideal leads affect
only the coefficients  $C_1$ and $C_2$ in Eq.53, which change
the absolute scale of $\rho$ and the origin of $n$.

\begin{center}
{\bf 6. Dependence of  $P(\rho)$ on the ideal leads}
\end{center}

According to (12) the elements of the transfer matrix $T$ obey
the relations $T_{22}=T^*_{11}$, $T_{21}=T^*_{12}$,
following from the time reversal invariance; so it is
sufficient to consider the elements
$$
T_{11}=x+iy\,, \qquad T_{12}=z+i\omega\,.
\eqno(58)
$$
If the mutual distribution $P(\rho, \varphi, \theta)$
is known for the parameters of the matrix (12), then the
distribution of $x,y,z,\omega$ is composed
according to the rule
$$
P_1(x,y,z,\omega) = \int \,d\rho \,d\varphi \,d\theta\,
P(\rho, \varphi, \theta) \, \cdot
$$
$$\qquad\qquad\cdot\,
\delta\left(x-\sqrt{1+\rho} \cos\varphi\right)
\delta\left(y-\sqrt{1+\rho} \sin\varphi\right) \cdot
$$
$$ \cdot\,
\delta\left(z-\sqrt{\rho} \cos\theta\right)
\delta\left(\omega-\sqrt{\rho} \sin\theta\right)\,.
\eqno(59)
$$
Setting $\rho=r^2$, $1\!+\!\rho=R^2$ and introducing
decomposition of unity under integration
$$
1= \int d R^2
\delta\left(R^2-r^2-1\right)\,,
\eqno(60)
$$
it is easy to find the relation between $P_1(x,y,z,\omega)$
and $P(\rho, \varphi, \theta)$:
$$
P_1(x,y,z,\omega) = P\left(z^2+\omega^2, {\rm Arctg}\frac{y}{x},
{\rm Arctg}\frac{\omega}{z}\right)\cdot
$$
$$\qquad\qquad\qquad\qquad
\cdot\, 4 \delta\left(x^2\!+\!y^2\!-\!z^2\!-\!\omega^2\!-\!1\right)\,.
\eqno(61)
$$
Inversely, if the distribution $P_1(x,y,z,\omega)$ is known,
then it always contains the delta function
$$
P_1(x,y,z,\omega) =  \tilde P(x,y,z,\omega)\,
 \cdot\, 4
\delta\left(x^2\!+\!y^2\!-\!z^2\!-\!\omega^2\!-\!1\right).
\eqno(62)
$$
and the distribution of $\rho$, $\varphi$, $\theta$ is
given by the relation
$$
P(\rho,\varphi, \theta) =
$$
$$=
\tilde P\left( \sqrt{1\!+\!\rho} \cos\varphi,\sqrt{1\!+\!\rho}
\sin\varphi, \sqrt{\rho} \cos\theta,\sqrt{\rho}
\sin\theta\right).
\eqno(63)
$$

A transformation of the matrix $T$ under the change of the ideal
leads is given by relations  (17), (18), which are rather
tremendous, but can be written in the compact form. As clear
from the stated above, the case $k=\kappa$ is special for a
situation in the forbidden band: it corresponds to the maximal
transparency of interfaces (see Eq.4), while the phase
$\psi=\theta-\varphi$ is localized near $\pi/2$ for all
$L$ (Sec.3). If the values of parameters for $k=\kappa$
are marked with a bar, then the relation (18) can be
rewritten in the form
$$
x\!=\!\bar x,\,\,\, y\!=\!\Delta_2\bar y\!-\!\Delta_1\bar \omega,\,\,\,
z\!=\!\bar z,\,\,\, \omega\!=\!-\Delta_1\bar y\!+\!\Delta_2\bar \omega,
\eqno(64)
$$
or inversely,
$$
\bar x\!=\!x,\,\,\, \bar y\!=\!\Delta_2 y\!+\!\Delta_1 \omega,\,\,\,
\bar z\!=\!z,\,\,\, \bar \omega\!=\!\Delta_1 y\!+\!\Delta_2 \omega,
\eqno(65)
$$
where $\Delta_1$ and $\Delta_2$ are defined in Eq.9. The
elements  $\bar T_{11}=\bar x+i\bar y$ and $\bar  T_{12}=\bar
z+i\bar \omega$ correspond to the momentum $\kappa$
and  are determined by the internal properties of the system,
while information on the ideal leads  (i.e. the Fermi
momentum $k$) contains in $\Delta_1$ and $\Delta_2$.

The case $k=\bar k$ is special for a situation in the allowed
band (Fig.2,a). Marking by a bar  the values of
parameters, corresponding to
this case and setting formally  $\bar k=\kappa$, one can
verify that the relation between  $T_{ij}$ and $\bar T_{ij}$
is also given by Eq.64,\,65. If the distribution function
$\bar P(\rho, \varphi, \theta)$ (determined by the
internal properties of the system) is known for two indicated
special situations, then (61) gives corresponding distribution
$\bar P_1(\bar x,\bar y,\bar z,\bar \omega)$, while the change
of variables (65) gives $P_1(x,y,z,\omega)$, and then (63)
gives the required distribution $P(\rho, \varphi,\theta)$,
depending on the ideal leads:
$$
P(\rho, \varphi, \theta)=
\bar P(\rho', \varphi', \theta')\,,
$$
$$
\rho'=\rho\cos^2\theta +\left(\Delta_1\sqrt{1\!+\!\rho}
\,\sin \varphi+\Delta_2\sqrt{\rho}\sin \theta\right)^2,
$$
$$
{\rm tg}\, \varphi'=
\frac{\Delta_2\sqrt{1\!+\!\rho}\,\sin \varphi+
\Delta_1\sqrt{\rho}\,\sin
\theta}{\sqrt{1\!+\!\rho}\,\cos\varphi}\,,
$$
$$
{\rm tg}\, \theta'=
\frac{\Delta_1\sqrt{1\!+\!\rho}\,\sin \varphi+
\Delta_2\sqrt{\rho}\,\sin
\theta}{\sqrt{\rho}\,\cos\theta} \,.
\eqno(66)
$$
In the forbidden band we have for $k=\kappa$
$$
\bar P(\rho, \varphi, \theta)= \bar P(\rho) \delta(\varphi)
\delta\left(\theta-\frac{\pi}{2}\right)\,,
\eqno(67)
$$
\onecolumn

\noindent
where the delta functions should be
broaden to the width
of  order $\epsilon$. If we neglect this widening, then
$$
P(\rho, \varphi, \theta)= \frac{1}{\Delta_2^2}
 \bar P\left(\frac{\rho}{\Delta_2^2}\right)
\delta\left(\theta-\frac{\pi}{2}\right)\cdot
 \delta\left(\varphi+{\rm arcsin}\frac{\Delta_1}{\Delta_2}
  \sqrt{\frac{\rho}{1\!+\!\rho}}\right)
\eqno(68)
$$
and integration over  $\varphi$ and  $\theta$ gives the result
(6), i.e. ideal leads do not change the form of $P(\rho)$ and
only renormalize the absolute scale of conductance.

In the deep of the allowed band for $k=\bar k$ (Fig.2,a)
the distribution $\bar P(\rho, \varphi, \theta)$  does not
depend on the phase variables\,\footnote{\,For this, strictly
speaking, one should average (32) over the $n$ variations
of order  $1/\delta$.  }
$$
\bar P(\rho, \varphi, \theta)=
\frac{1}{(2\pi)^2} \bar P(\rho) = \frac{1}{(2\pi)^2\,t}
\exp\left(-\frac{\rho}{t}\right)
\,,
\eqno(69)
$$
and $\bar P(\rho)$ is determined by Eq.7 with $\gamma=0$.
For the foreign leads (Fig.2,b) the distribution of
$\rho$ is given by the integral
$$
P(\rho)=\frac{1}{(2\pi)^2\,t} \int\limits^{\pi}_{-\pi} d\varphi
\int\limits^{\pi}_{-\pi} d\theta
\exp\left\{-\frac{\rho+S(\rho,\varphi,\theta)}{t}\right\}
\,,
\eqno(70)
$$
where
$$
S(\rho,\varphi,\theta) =
\Delta^2_1(1\!+\!\rho) \,\sin^2 \varphi+
2\Delta_1 \Delta_2 \sqrt{(1\!+\!\rho)\rho} \,\sin \varphi
\sin \theta +
\Delta_1^2\rho\sin^2 \theta  \,.
\eqno(71)
$$
Calculation of the integral (see Appendix 2) gives the following
results.

In the case $\Delta^2_1\ll 1$, two regions are actual,
$t\ll \Delta^2_1$ and $t\gg \Delta^2_1$; in the first of
them
$$
P(\rho)=\left \{ \begin{array}{cccc}
\displaystyle{ \sqrt{\frac{1}{\pi\Delta^2_1 t}}
\exp\left(-\frac{\rho}{2t} \right) }\,,&  \rho\alt t \\
{  }& {  } \\
\displaystyle{
\frac{1}{\pi} \sqrt{\frac{1}{\rho(\rho_c\!-\!\rho)}}  }
\,,& t\alt \rho\,, \quad \rho_c\!-\!\rho \,\agt
\,\left(t\Delta^2_1\right)^{1/2}
\\ {  }& {  } \\ \displaystyle{ \frac{1}{2\pi^2}
\sqrt{\frac{\pi}{2\rho \left(t\Delta^2_1\right)^{1/2} }}\,
\Gamma(1/4) }\,, &
|\rho_c\!-\!\rho| \alt\, \left(t\Delta^2_1\right)^{1/2} \\
\ {  }& {  } \\
\displaystyle{
\frac{1}{\pi\sqrt{AB}}  \exp\left(-\frac{S_c}{t}
\right) } \,,& \rho\!-\!\rho_c \agt \,
\left(t\Delta^2_1\right)^{1/2}
\end{array} \right.\,,
\eqno(72)
$$
where
$$
S_c=\left(\Delta_1 \sqrt{1\!+\!\rho} -\Delta_2 \sqrt{\rho}
\right)^2 \Theta(\rho-\rho_c) \,, \qquad
\rho_c= \Delta_1^2 \,,
$$
$$
A= 2 \Delta_1 \sqrt{1\!+\!\rho} \left(\Delta_2
\sqrt{\rho} -\Delta_1 \sqrt{1\!+\!\rho} \right) \,, \quad
B=2 \Delta_1 \sqrt{\rho}\left(\Delta_2 \sqrt{1\!+\!\rho} -\Delta_1
\sqrt{\rho} \right) \,,
\eqno(73)
$$
and the main probability corresponds to the second result in (72).
In the region $t\gg \Delta^2_1$ one has
$$
P(\rho)=\left \{ \begin{array}{cc}
\displaystyle{ \frac{1}{t} \exp\left(-\frac{\rho}{t} \right)
\left[1-\frac{\Delta^2_1}{2t}-\frac{\Delta^2_1 \rho}{t}
+\frac{\Delta^2_1 \Delta^2_2 \rho(1\!+\!\rho)}{2t^2} \right]
 }\,,&  \rho(1\!+\!\rho)\alt t^2/\Delta_1^2 \\
{  }& {  } \\
\displaystyle{
\frac{1}{\pi\sqrt{AB}} \exp\left(-\frac{S_c}{t}
\right) } \,,&  \rho(1\!+\!\rho)\,\agt \,t^2/\Delta_1^2
\end{array} \right.\,,
\eqno(74)
$$
and the normalization integral is determined by the first
asymptotics. Evolution of the distribution for $\Delta_1^2\ll 1$
is shown in Fig.5,a. At small $t$ it reduces to the smearing
of singularities of the distribution (8), while in the limit
of large $t$ the value $P(\rho)$ at $\rho=0$ tends to one
of the distribution $\bar P(\rho)$ and the whole form of
the distribution is close to the latter.

In the case $\Delta_1^2\gg 1$, three regions are actual,
$t\ll 1$, $1\ll t\ll \Delta_1^2$ and $t\gg \Delta_1^2$.
In the first of them one has results
$$
P(\rho)=\left \{ \begin{array}{cccc}
\displaystyle{ \sqrt{\frac{1}{\pi\Delta^2_1 t}}
\exp\left(-\frac{\rho}{2t} \right) }\,,&  \rho\alt t \\
{  }& {  } \\
\displaystyle{
\frac{1}{\pi} \sqrt{\frac{1}{\rho(\rho_c\!-\!\rho)}}  }
\,,& t\alt \rho\,, \quad \rho_c\!-\!\rho \agt \,
\left(t\Delta^2_1\rho\right)^{1/2} \\
{  }& {  } \\
\displaystyle{
\frac{1}{2\pi^2} \sqrt{\frac{\pi}
{2\rho \left(t\Delta^2_1\rho\right)^{1/2}}}
\,\Gamma(1/4) }
\,,&  |\rho_c\!-\!\rho| \alt \,
\left(t\Delta^2_1\rho\right)^{1/2} \\
{  }& {  } \\
\displaystyle{
\frac{1}{\pi\sqrt{AB}} \exp\left(-\frac{S_c}{t}
\right) } \,,& \rho\!-\!\rho_c \agt\,
\left(t\Delta^2_1\rho\right)^{1/2}
\end{array} \right.\,,
\eqno(75)
$$
which are analogous to (72) and correspond to smearing of
singularities of the distribution (8); the main probability
corresponds to the second asymptotics. In the region
$1\ll t\ll \Delta_1^2$ we obtain
$$
P(\rho)=\left \{ \begin{array}{ccccc}
\displaystyle{  \sqrt{\frac{1}{\pi\Delta^2_1 t}}
\exp\left(-\frac{\rho}{2t} \right) }\,,&  \rho\alt 1 \\
{  }& {  } \\
\displaystyle{
\frac{1}{\pi^2} \sqrt{\frac{\pi}{\rho\Delta^2_1 t}}\ln \rho }
\,,& 1\alt\rho\alt  t  \\
{  }& {  } \\
\displaystyle{
\frac{1}{\pi^2} \sqrt{\frac{\pi}{\rho\Delta^2_1 t}}\ln t }
\,,& t\alt\rho\alt \Delta^2_1/t   \\
{  }& {  } \\
\displaystyle{
\frac{1}{2\pi^2} \sqrt{\frac{\pi}{\rho\Delta^2_1 t}}
\ln \frac{\Delta^2_1 t}{\rho}
\exp\left(-\frac{S_c}{t} \right) }
\,,& \Delta^2_1/t \alt\rho\alt \Delta^2_1 t   \\
{  }& {  } \\
\displaystyle{
\frac{1}{\pi\sqrt{AB}} \exp\left(-\frac{S_c}{t}
\right) } \,,&  \rho\agt \Delta^2_1 t
\end{array} \right.
\eqno(76)
$$
while in the region $t\gg \Delta_1^2$
$$
P(\rho)=\left \{ \begin{array}{cccc}
\displaystyle{ \frac{1}{t} \exp\left(-\frac{\rho}{t} \right)
\left[1-\frac{\Delta^2_1}{2t}-\frac{\Delta^2_1 \rho}{t}
+\frac{\Delta^2_1 \Delta^2_2 \rho(1\!+\!\rho)}{2t^2} \right]
 }\,,&  \rho\alt t/\Delta_1^2 \\
{  }& {  } \\
\displaystyle{
\frac{1}{\pi^2} \sqrt{\frac{\pi}{\rho\Delta^2_1 t}}
\ln \frac{\Delta^2_1 \rho}{t} }
\,,& t/\Delta^2_1 \alt\rho\alt  t   \\
{  }& {  } \\
\displaystyle{
\frac{1}{2\pi^2} \sqrt{\frac{\pi}{\rho\Delta^2_1 t}}
\ln \frac{\Delta^2_1 t}{\rho}
\exp\left(-\frac{S_c}{t} \right) }
\,,& t \alt\rho\alt \Delta^2_1 t   \\
{  }& {  } \\
\displaystyle{
\frac{1}{\pi\sqrt{AB}} \exp\left(-\frac{S_c}{t}
\right) }  \,,&  \rho\agt \Delta_1^2 t
\end{array} \right.
\eqno(77)
$$
In both cases the main probability is related with two last
asymptotic results. The characteristic feature of (76)
and (77) is existence of the quick exponent ($\exp(-\rho/2t)$
or $\exp(-\rho/t)$) for small $\rho$  and the slow
\twocolumn

\noindent
exponent
$\exp(-\rho/\Delta_1^2 t)$ for large $\rho$, while the
power law behavior $P(\rho)\propto \rho^{-1/2}$ is valid for
intermediate $\rho$, apart of the logarithmic corrections.
Evolution of the distribution for $\Delta_1^2\gg 1$ is shown in
Fig.5,b.

In the above discussion we had in mind that results depend
on $|\Delta_1|$ (which can be tested by the change
$\varphi\to\varphi+\pi$ in Eqs.70,\,71) and $\Delta_1$
can be consider as positive without the loss of generality.
In addition, we assumed that the distribution (69) is given
axiomatically and did not discuss its conditions of
applicability: it allows to understand better, how
$P(\rho)$ is transformed due to ideal leads.
In fact, the distribution (69) corresponds to solution of
equation (2) only for $\rho\alt 1$, which restricts
the physical actuality of  (76), (77) by the condition
$\rho\alt \Delta_1^2$.

In the log-normal regime, arising for $t\gg 1$, one should
take $\bar P(\rho)$  in the form (5) with $\gamma=0$.
Since the typical values of $\rho$ are large, we can
set $\rho'=\rho K(\varphi,\theta)$ in Eq.66, where
$$
K(\varphi,\theta)=\cos^2\theta +\left(\Delta_1
\,\sin \varphi+\Delta_2\sin \theta\right)^2\,.
\eqno(78)
$$
Substitution to (5), (66), (69) gives
$$
P(\rho,\varphi,\theta)= \frac{1}{(2\pi)^2}
\frac{1}{\rho K \sqrt{4\pi t}}
\exp\left\{-\frac{(\ln K\rho-t)^2}{4t}\right\} \approx
$$
$$
\approx
\frac{1}{(2\pi)^2}
\frac{1}{\rho \sqrt{4\pi t}}
\exp\left\{-\frac{(\ln \rho-t)^2}{4t}\right\}\cdot
$$
$$ \qquad\qquad\qquad\qquad\cdot
\left[\frac{1}{K}- \frac{2(\ln \rho-t) \ln K}{4t K}
 \right] \,,
\eqno(79)
$$
where expansion in $1/t$ is produced, using the fact
that $(\ln \rho-t)\sim\sqrt{t}$ for the bulk of the
distribution. Integrating over $\varphi$ and $\theta$,
we set
$$
\int\limits^{\pi}_{-\pi} \frac{d\varphi}{2\pi}
\int\limits^{\pi}_{-\pi} \frac{d\theta }{2\pi}  \,
\frac{1}{K(\varphi,\theta)} \equiv \frac{1}{K_0}\,,\qquad
$$
$$
\int\limits^{\pi}_{-\pi} \frac{d\varphi}{2\pi}
\int\limits^{\pi}_{-\pi} \frac{d\theta }{2\pi} \,
\frac{\ln K(\varphi,\theta)}{K(\varphi,\theta)}
\equiv \frac{\ln K_1}{K_0}\,.
\eqno(80)
$$
Then conservation of probability requires the condition
$K_0=1$, under which $P(\rho)$ can be written in the
form
$$
P(\rho)= \frac{1}{(2\pi)^2}
\frac{1}{\rho \sqrt{4\pi t}}
\exp\left\{-\frac{(\ln K_1\rho-t)^2}{4t}\right\} \,.
\eqno(81)
$$
The equality $K_0=1$ is indeed valid, as one can verify by
a direct calculation of the integral. We see that the ideal
leads do not change the parameters of the log-normal distribution
and only renormalize the absolute scale of $\rho$, which
is determined by the parameter $K_1$ (Fig.8):
$$
K_1=\left \{ \begin{array}{cc}
\displaystyle{ 1-\Delta^2_1/2 }\,,&  \Delta_1\ll 1 \\
{ {\rm const}/ \Delta^2_1\,, }& { \Delta^2_1\gg 1 }
\end{array} \right.\,,
\eqno(82)
$$
where the constant is numerically close to 4.

\begin{figure}
\centerline{\includegraphics[width=2.8 in]{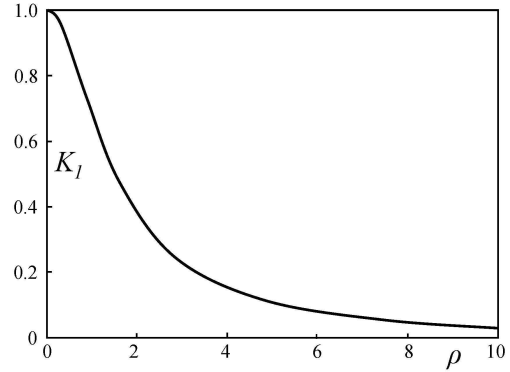}}
\caption{Parameter $K_1$ as function of $\Delta_1$.
} \label{fig8}
\end{figure}

In the critical  region for sufficiently small $n$ we can
begin with Eq.38, which determines the distribution
$\bar P(\rho,\varphi,\theta)$ in the form (39).
A transformation of variables (64) results in the change
$S\to S(\Delta_2\!-\!\Delta_1)$ in (38),(39) and
renormalization (6) of the absolute scale of conductance
with $A=1/(\Delta_2\!-\!\Delta_1)^2$.

\begin{center}
{\bf 7. Evolution equation for $P(\rho)$ in the presence of the
ideal leads }
\end{center}

According to Sec.3, the transfer matrix $T$ for a succession of
point scatterers and the "natural" ideal leads (Fig.2,a)
has a form (22); the corresponding parameters $\epsilon_s$
will be denoted as $\bar\epsilon_s$. In the presence of foreign
leads (Fig.2,b) the system is described by the transfer
matrix $\tilde T=T_a T T_b$. Inserting  the product $ T_b
T_a=1$ between each two multipliers of (22), one has
$$
\tilde T^{(n)}=\tilde T_{\epsilon_0}
\,\tilde T_{\delta_1} \, \tilde T_{\epsilon_1}\,
\tilde T_{\delta_2}\,\tilde T_{\epsilon_2}\,
\ldots\,
\tilde T_{\delta_n}\,\tilde T_{\epsilon_n}\,,
\eqno(83)
$$
where
$$
\tilde T_{\epsilon_s}=  T_a \,T_{\epsilon_s}\,  T_b \,,\qquad
\tilde T_{\delta}=  T_a \,T_{\delta}\,  T_b\,.
\eqno(84)
$$
In the allowed band the difference between $\tilde T_{\epsilon_s}$
and $ T_{\epsilon_s}$ is reduced to renormalization
$\bar\epsilon_s \to \epsilon_s =\bar\epsilon_s \bar k /k$,
which has a simple physical sense. Let represent the initial
system (Fig.9,a) as a succession of similar scatterers
(Fig.9,b), creating the potential well for each point
scatterer with the width tending to zero. Then the parameter
$\bar\epsilon_s=V_s/2\bar k a_0$, defined for the
Fermi momentum $\bar k$ of the system, is replaced by the
parameter $\epsilon_s=V_s/2k a_0$, containing the Fermi momentum
$k$ in the ideal leads.
As for $\tilde T_{\delta}$, it is the
transfer matrix of the potential barrier, separating two point
scatterers in Fig.9,b:
$$
\tilde T_{\delta}= \,\left ( \begin{array}{cc}
\cos\delta -i\Delta_2\sin\delta & i\Delta_1\sin\delta \\
-i\Delta_1\sin\delta & \cos\delta +i\Delta_2\sin\delta
\end{array} \right)\,
\approx
$$
$$
\approx
\,\left ( \begin{array}{cc}
1-i\Delta_2\delta & i\Delta_1\delta \\
-i\Delta_1\delta & 1+i\Delta_2\delta
\end{array} \right)\,,
\eqno(85)
$$
where $\Delta_1$ and $\Delta_2$ are determined by
Eq.9, if one set $\bar k=\kappa$. The situation in the forbidden
band (Fig.2,c) differs only by the fact, that the height of
barriers in Fig.9,b prevails the Fermi level, while $\tilde
T_{\delta}$ is obtained from Eq.85 by the replacement of
$\bar k$ by $i\kappa$
$$
\tilde T_{\delta}=\,\left ( \begin{array}{cc}
{\rm ch}\delta -i\Delta_1{\rm sh}\delta & i\Delta_2{\rm sh}\delta
\\ -i\Delta_2{\rm sh}\delta
& {\rm ch}\delta +i\Delta_1{\rm sh}\delta
\end{array} \right)\,
\approx
$$
$$
\approx
\,\left ( \begin{array}{cc}
1-i\Delta_1\delta & i\Delta_2\delta \\
-i\Delta_2\delta & 1+i\Delta_1\delta
\end{array} \right)   \,.
\eqno(86)
$$
Let describe the potential barrier by the transfer matrix of
the general form
$$
\tilde T_{\delta}= \left ( \begin{array}{cc}
\!\!\! A &\!\! \!\!B \!\!\\ \!B^* & \!\!A^* \!\!\end{array} \right)=
\left ( \begin{array}{cc}
\sqrt{1\!+\!\Delta^2} {\rm e}^{i\alpha} & \!\!\Delta {\rm e}^{-i\beta}\!\!
\\ \Delta {\rm e}^{i\beta} & \!\!\sqrt{1\!+\!\Delta^2} {\rm e}^{-i\alpha}\!\!
\end{array} \right),
\eqno(87)
$$
close to the unit one ($\alpha,\,\Delta\ll
1$); it has eigenvalues ${\rm e}^{\pm \delta}$, where
$\delta^2=\Delta^2-\alpha^2$. According to (83),
$\tilde T^{(n)}$ can be
expressed
as $\tilde T^{(n-1)}$
multiplied by $\tilde T_{\delta} \tilde T_{\epsilon_n}$.
Accepting $\tilde T^{(n-1)}$ in the form  (12), one has
$$
\tilde T^{(n)}_{12}= \sqrt{1\!\!+\!\!\rho}\, {\rm e}^{i\varphi}
(B\!+\!i\epsilon C) +\sqrt{\rho}\, {\rm e}^{i\theta} (A^*\!\!-\!i\epsilon
C^*) \,,
\eqno(88)
$$
where $C=B-A$. Squaring the modulus of (88), we have the
value of $\tilde \rho$, corresponding to $\tilde T^{(n)}$
$$
\tilde\rho= \rho+D \sqrt{\rho(1\!+\!\rho)}  +\epsilon^2
(1\!+\!2\rho)\,,
 \eqno(89)
 $$
where
$$
D= 2\Delta \cos{(\psi\!+\!\beta)} -2\epsilon \sin{\psi}
-2\epsilon^2  \cos{\psi}\,,
\eqno(90)
$$
$$
\psi=\theta-\varphi\,,
\eqno(91)
$$
and we  retained the terms of the first order in $\delta$ and
the second order in $\epsilon$. Expression (89) is analogous
to Eq.$(A.2)$ of the paper  \cite{12} and subsequent calculations
follow to Appendix $A$ of this paper.  As a result, we obtain
Eq.3 with parameters (we set $a_0=1$)
$$
\tilde\alpha=\frac{1}{2} \overline{D^2}\,,\quad
\gamma_1\tilde\alpha=\epsilon^2-\frac{1}{2}
\overline{D^2}\,,\quad
\gamma_2\tilde\alpha=\frac{1}{2}\overline{D}\,,
\eqno(92)
$$
and substitution of (90) gives
$$
\tilde\alpha=2\epsilon^2 \overline{\sin^2{\psi}}\,,\qquad
\gamma_1\tilde\alpha=\epsilon^2\left(1- 2
\overline{\sin^2{\psi}}\right)
\,,
$$
$$
\gamma_2\tilde\alpha=\Delta\overline{\cos{(\psi\!+\!\beta)}}
-\epsilon^2\overline{\cos{\psi}}\,,
\eqno(93)
$$
where $\beta=-\pi/2$ in the actual case; another values of
$\beta$ occur in the presence of the $\delta$-function
potential on interfaces.

\begin{figure}
\centerline{\includegraphics[width=2.5 in]{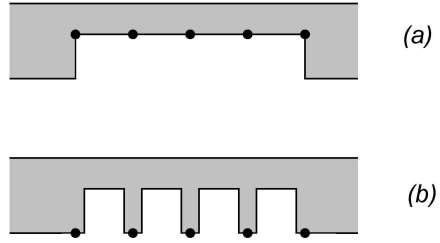}}
\caption{\footnotesize The physical sense of the $\epsilon_s$
renormalization  consists in representation of the initial
system (a) as a succession of  similar scatterers
(b).  In the latter case the potential well is created
for each point scatterer, whose width tends to zero.
} \label{fig9}
\end{figure}

Equation (3) with parameters (93) allows to analyze the
special situations with $\epsilon=\bar\epsilon$, actual
for Sec.6. In the deep of the allowed band and $k=\bar k$
we have $\Delta_1=0$ and $\Delta=0$ (see (85),(87)); in the
case of the uniform distribution for $\psi$, Eq.93 gives
$\tilde\alpha =\epsilon^2$, $\gamma_1=0$, $\gamma_2=0$,
in agreement with the results of Secs.4,\,5. Hence, the uniform
distribution for $\psi$ obtained in Sec.3 for $n\alt 1/\epsilon^2$
retains also for $n\agt 1/\epsilon^2$. In the deep of the
forbidden band and for $k=\kappa$ one has $\Delta_1=0$ and
$\Delta=\delta$ (see (86),(87)); the results
$\tilde\alpha=2\epsilon^2$, $\gamma\tilde\alpha=\delta$,
obtained in Secs.4,\,5, agree with (93), if localization of
$\psi$ near $\pi/2$ is accepted; such localization was
established in Sec.3 for $n\alt 1/\epsilon^2$ and retains for
$n\agt 1/\epsilon^2$. In the critical region
$\Delta_1\approx \Delta_2$ and  $\Delta\sim \delta$, so the
results $\tilde\alpha\sim \left(\epsilon^2\delta^2\right)^{1/3}$,
$\gamma=\gamma_1+\gamma_2\sim 1$ of Sec.4,\,5 agree with
(93) in the case of localization of $\psi$ in the region of
small values $\sim\left(\delta/\epsilon^2\right)^{1/3}$, in
accordance with the analysis of Sec.3.

\begin{center}
{\bf 8. Evolution of the mean value of $\rho$.}
\end{center}

\onecolumn

Evolution of $\langle\rho\rangle$ allows a complete
description and illustrates the influence of
the ideal leads for the
arbitrary system length $L$.
The relation $\tilde T^{(n)}=\tilde T^{(n-1)}\tilde T_{\delta}
\tilde T_{\epsilon_n}$ has the following form in terms of the
matrix elements
$$
\tilde T_{11}^{(n)}=
(1+i\alpha-i\epsilon_n)\, \tilde T_{11}^{(n-1)} +
(-i\Delta+i\epsilon_n) \,\tilde T_{12}^{(n-1)}\,,\quad
\tilde T_{12}^{(n)}=
(i\Delta-i\epsilon_n) \, \tilde T_{11}^{(n-1)} +
 (1-i\alpha+i\epsilon_n)\,\tilde T_{12}^{(n-1)}  \,,
\eqno(94)
$$
where
we linearized in $\alpha$, $\Delta$
and set $\beta=-\pi/2$. Introducing notations for the second
moments
$$
z_{1}^{(n)}=\left\langle
\left|\tilde T_{11}^{(n)}\right|^2\right\rangle\,, \qquad
z_{2}^{(n)}=\left\langle
\tilde T_{11}^{(n)} \tilde T_{12}^{(n)*}\right\rangle\,, \qquad
z_{3}^{(n)}=\left\langle
\tilde T_{11}^{(n) *} \tilde T_{12}^{(n)}\right\rangle\,, \qquad
z_{4}^{(n)}=\left\langle
\left|\tilde T_{12}^{(n)}\right|^2\right\rangle\,, \qquad
 \eqno(95)
$$
we have the system of difference equations
$$
\left ( \begin{array}{cccc} z_{1}^{(n)} \\ z_{2}^{(n)}
\\ z_{3}^{(n)} \\ z_{4}^{(n)}\end{array} \right)\,
= \left ( \begin{array}{cccc} 1+\epsilon^2 &
i\Delta -\epsilon^2 & -i\Delta -\epsilon^2 &\epsilon^2\\
-i\Delta -\epsilon^2 & 1+2i\alpha -\epsilon^2 &
-\epsilon^2 & -i\Delta +\epsilon^2 \\
i\Delta -\epsilon^2 & -\epsilon^2  &1-2i\alpha -\epsilon^2
 & i\Delta +\epsilon^2 \\
\epsilon^2 &
i\Delta -\epsilon^2 & -i\Delta -\epsilon^2 &1+\epsilon^2
 \end{array} \right)\,
\left ( \begin{array}{cccc} z_{1}^{(n-1)} \\ z_{2}^{(n-1)}
\\z_{3}^{(n-1)} \\z_{4}^{(n-1)} \end{array} \right) \,,
\eqno(96)
$$
whose solution is exponential, $z_{i}^{(n)} \sim \lambda^n$,
and $\lambda$ is an eigenvalue of the matrix; substitution
$\lambda=1+x$ leads to the equation
$$
x\left[
x\left(x^2+4\alpha^2-4\Delta^2\right)- 8\epsilon^2
(\Delta-\alpha)^2 \right] =0  \,,
\eqno(97)
$$
which has the solution $x=0$ and three nontrivial roots.
The latter coincide with roots of Eq.44 and do not
depend on the ideal leads, since $\delta^2=\Delta^2-\alpha^2$
and
$$
8\epsilon^2 (\Delta-\alpha)^2 = 8\epsilon^2 \delta^2
\left(\Delta_2-\Delta_1\right)^2 =
8\bar\epsilon^2 \delta^2 =2 W^2  \,.
\eqno(98)
$$
Finding the eigenvectors of matrix (96), we have the
general solution for $z_i^{(n)}$
$$
\left ( \begin{array}{cccc} z_{1}^{(n)} \\ z_{2}^{(n)}
\\ z_{3}^{(n)} \\ z_{4}^{(n)}\end{array} \right)\,
=C_0\,\left ( \begin{array}{cccc} -1\\ 0\\ 0 \\ 1
\end{array} \right)\,
+\sum\limits_{i=1}^3\, C_i  \,
\left ( \begin{array}{cccc} 1 \\ e_{2}(x_i)
\\ e_{3}(x_i)\\ 1 \end{array} \right)\, \exp{(x_i n)} \,,
\eqno(99)
$$
where $x_1,\,\,x_2,\,\,x_3$ are nontrivial roots of (97),
and
$$
e_{2}(x)=\frac{{\cal A}x+{\cal B}}{p(x)}\,,\qquad
e_{3}(x)=\frac{{\cal A}^* x+{\cal B}^*}{p(x)}\,,
$$
$$
{\cal A}=2\epsilon^2 -2i\Delta\,,
\qquad {\cal B}=4\alpha\Delta + 4i\epsilon^2 (\alpha-\Delta)\,,
\eqno(100)
$$
$$
p(x)=x^2+2\epsilon^2 x +4 \alpha^2 \,.
$$
Choosing the unit transfer matrix as the initial condition, we
have  $z_1^{(0)}=1$, $z_2^{(0)}=z_3^{(0)}=z_4^{(0)}=0$, which
allows to establish the coefficients $C_i$. Since $z_4^{(n)}$
gives immediately $\langle \rho\rangle$, then
$$
\langle \rho\rangle = -\frac{1}{2}
+\frac{(x_2\!-\!x_3)\,p(x_1)}{2Q}\, {\rm e}^{x_1 n}
-\frac{(x_1\!-\!x_3)\,p(x_2)}{2Q}\, {\rm e}^{x_2 n}
+\frac{(x_1\!-\!x_2)\,p(x_3)}{2Q}\, {\rm e}^{x_3 n} \,,
\eqno(101)$$
where
$$
Q = x_1^2\, (x_2\!-\!x_3) - x_2^2\, (x_1\!-\!x_3) +
x_3^2\, (x_1\!-\!x_2) \,.
\eqno(102)
$$
The result (101) has the general character, since no
approximations were made in its derivation. For large
$n$ one of the exponents is dominated, and a situation is
multiplicative (Sec.1). For small $n$ the exponents
can be extended in the series, recovering the Ohmic
regime $\rho\propto n$.
\twocolumn

In the localized regime ($\delta\gg \bar\epsilon^2$),
the nontrivial roots of Eq.97 have a form
$$
x_1= 2\delta+\bar\epsilon^2, \quad
x_2= -2\bar\epsilon^2, \quad
x_3= -2\delta+\bar\epsilon^2,
\eqno(103)
$$
and
$$
\langle \rho\rangle = -\frac{1}{2}
+\frac{\Delta_2^2}{4}\,\left( {\rm e}^{x_1 n}+
{\rm e}^{x_3 n} \vphantom{e^2_5} \right)
-\frac{\Delta_1^2}{2}\, {\rm e}^{x_2 n}  \,.
\eqno(104)
$$
For large $n$ the term ${\rm e}^{x_1 n}$ is dominated,
confirming renormalization of $\rho$ by the factor
$\Delta_2^2$, indicated in Eq.6; the change of the
$n$ origin  (see the end of Sec.5) is essential only for $n\alt
1/\delta$.

In the metallic regime one makes the change
$\delta\to i\delta$, $\epsilon\to -i\epsilon$ and considers
the limit $\delta\gg\bar\epsilon^2$; then
$$
x_1= 2\bar\epsilon^2, \quad
x_2=\! -\bar\epsilon^2\!+\!2i\delta\,, \quad
x_3=\! -\bar\epsilon^2\!-\!2i\delta\,,
\eqno(105)
$$
and
$$
\langle \rho\rangle = -\frac{1}{2}
+\frac{\Delta_2^2}{2}\, {\rm e}^{2\bar\epsilon^2 n}
-\frac{\Delta_1^2}{2}\, {\rm e}^{-\bar\epsilon^2 n}
\cos{2n\delta}
 \,,
\eqno(106)
$$
which for the "natural" leads (when $\Delta_1=0$) gives the
well-known result \cite{1,2,15,16,17}
$$
\langle \rho\rangle = \frac{1}{2}
\left( {\rm e}^{2\bar\epsilon^2 n} -1 \right) \,.
\eqno(107)
$$
For finite $\Delta_1$, the oscillations arise whose period is
determined by the de Broglie wavelength (since $2n\delta=2\bar k
L$). Their origin is clear, since for $\bar\epsilon=0$
the Landauer resistance $\rho$ is determined by the
transfer matrix (85) of the potential barrier,
$$
\rho = \Delta_1^2 \sin^2{\bar k L}  \,,
\eqno(108)
$$
which becomes transparent,
if the system length
corresponds to the  semi-integer number of the de Broglie
wavelengths (analogously to blooming in optics). For finite
$\bar\epsilon$, the oscillations become attenuating, but
remain completely observable: the mean value of $\rho$ is
representative in the metallic regime, since its
fluctuations are relatively small.

In the critical region ($\delta\ll \bar\epsilon^2$) one
has
$$
x_1= 2\left(\bar\epsilon^2 \delta^2\right)^{1/3}, \quad
x_2= x_1 {\rm e}^{2\pi i/3}, \quad
x_3= x_1 {\rm e}^{-2\pi i/3},
\eqno(109)
$$
and
$$
\langle \rho\rangle = \frac{1}{6}
\left[ {\rm e}^{x_1 n} + 2 {\rm e}^{-x_1 n/2}
\cos{\left( \frac{\sqrt{3}}{2} x_1 n\right)} -3 \right] \,+
$$
$$
+\frac{\epsilon^2}{3 x_1}
\left[ {\rm e}^{x_1 n} - 2 {\rm e}^{-x_1 n/2}
\cos{\left( \frac{\sqrt{3}}{2} x_1 n+\frac{\pi}{3}\right)}
\right]+
\eqno(110)
$$
$$
+\frac{2\alpha^2}{3 x_1^2}
\left[ {\rm e}^{x_1 n} - 2 {\rm e}^{-x_1 n/2}
\cos{\left( \frac{\sqrt{3}}{2} x_1 n-\frac{\pi}{3}\right)}
\right] \,.
$$
If the critical region is approached from the allowed band,
then the "natural" ideal leads can be used (Fig.2,a), when
$\epsilon=\bar\epsilon$ and the second term is
dominated in Eq.110. Expansion in $x_1 n$ gives
$\langle \rho\rangle=\epsilon^2 n$ in agreement with the
transfer matrix (38), and establishes its range of
applicability as $n\alt 1/x_1$.

The result (110) remains finite in the limit $\bar k\to 0$,
when
$$
\frac{\epsilon^2}{3 x_1}=
\frac{\left(2W^2\right)^{2/3}}{24 k^2 a_0^2} \,, \qquad
\frac{2\alpha^2}{3 x_1^2} =
\frac{2k^2 a_0^2}{3\left(2W^2\right)^{2/3}} \,,
$$
$$
\qquad x_1=\left(2W^2\right)^{1/3} \,.
\eqno(111)
$$
Analogously to (106), the attenuating oscillations take place,
whose amplitude depends essentially on the Fermi momentum
$k$ in the ideal leads: it is of the order of unity for
$k a_0\sim W^{2/3}$ (when three terms in Eq.110 are
of the same order), but increases for small $k$ (when
the second term is dominated), as well as for large $k$
(when the third term is prevailed). The period of the
oscillations is determined by the amplitude of the random
potential, while the phase shift changes from $\pi/3$ till
$-\pi/3$ when $k$ is increased.

In fact, the observable picture is more complicated. In
the critical region fluctuations of $\rho$ are large, and
the form of the distribution function essentially depends
on the several first moments\,\footnote{\,Let us
remind that the Fourier transform of $P(\rho)$ gives
the characteristic function
$F(t)=\left\langle e^{i\rho t} \right\rangle$, which is
the generating function of moments,
$F(t)=\sum_{n=0}^{\infty}(it)^n\left\langle \rho^n
\right\rangle/n!$. If all moments of the distribution are known,
then one can construct $F(t)$, while $P(\rho)$ is given by the
inverse Fourier transform.  }.
Meanwhile, the higher moments
of $\rho$ are also oscillating with the period of the
same order: it is related with the complex roots of Eq.48
and analogous equations for higher moments
\,\footnote{\,Comparison of (44) and (48) shows that the periods
of oscillations for  $\langle\rho\rangle$ and
$\langle\rho^2\rangle$ differ by a factor $21^{1/3}$.
As clear from Sec.4 (see Footnote 3), the right hand side
of the equation for $x$ may contain only combinations
$\delta^{2n}\epsilon^{2m}$ with $n\ge m$,
of which only $\delta^{2n}\epsilon^{2n}\sim W^{2n}$
remain finite in the $\delta\to 0$ limit. Since for
$x\sim \delta\sim\epsilon^2$ all terms of the equation
have the same order of magnitude, the exponent of growth
$x$ for $\langle \rho^n \rangle$
at $\delta=0$ satisfies the equation $x^{2n+1}= c_1
\delta^{2}\epsilon^{2} x^{2n-2} + c_2 \delta^{4}\epsilon^{4}
x^{2n-5}+\ldots\,$, whose nontrivial roots are of order
$\left(\delta^{2}\epsilon^{2}\right)^{1/3}$
independently of $n$.}.
As a result, the complicated interference of incommensurate
oscillations occurs, so the resistance $\rho$ of the specific
sample undergoes aperiodic oscillations. Such oscillations were
observed in the magneto-resistance of thin wires \cite{36}
(Fig.10) and have the close relation to the above discussion:
the magnetic field, perpendicular to a wire, creates the quadratic
potential along it, which effectively restricts the length
of the system; so variation of the magnetic field is analogous
to the change of $L$. In principle, Fig.10 is explained by the
theoretical results of the papers \cite{25,26}, but we
are unaware on attempts of description of the oscillations
themselves.

\begin{figure}
\centerline{\includegraphics[width=2.8 in]{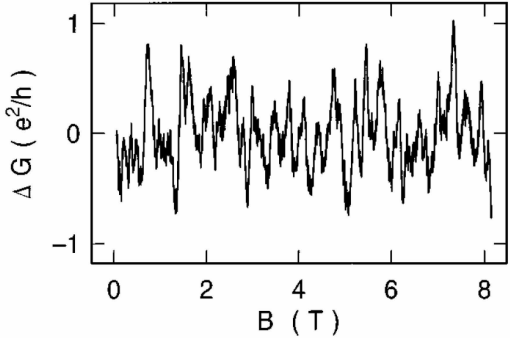}}
\caption{\footnotesize
The conductance $G(B)$ of the thin Au wire  \cite{36}
in the magnetic
field $B$ undergoes the "universal" fluctuations of order
$e^2/h$ due to a change of the impurity configuration \cite{25,26}.
Fluctuations of  $G(B)$  and $G(B+\Delta B)$ are
statistically independent, if $\Delta B$ exceeds a
certain characteristic scale \cite{25,26}. In spite of the
random character, the fluctuation picture is completely
reproducible and reflects the specific realization of the
random potential ("magnetic fingerprints").
} \label{fig10}
\end{figure}

Fig.11 shows the experimental results of the paper \cite{37},
which demonstrate a possibility of observation of the entire
distribution function of $\rho$ or $G$, as well as its
moments. Independent impurity configurations in the thin wire
of Si-doped GaAs  were created by the periodic
warming till the room temperature, and then averaging over
50 configurations were made. The results for the first two
moments of $G$ indicate that the distribution function
is not stationary, but undergoes systematic variations,
in agreement with the above arguments.

\begin{figure}
\centerline{\includegraphics[width=2.8 in]{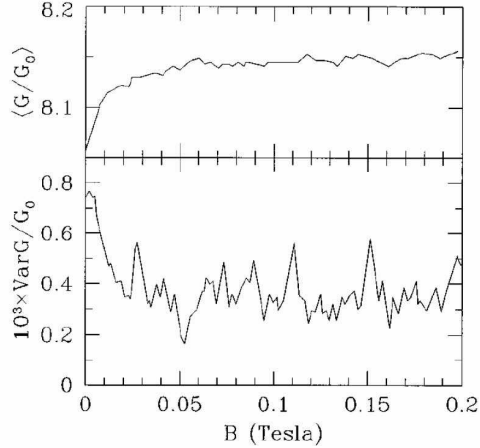}}
\caption{\footnotesize
Evolution of the first two moments of
the conductance $G$ of the thin Si-doped GaAs wire
under the change of the magnetic field
\cite{37}.} \label{fig11}
\end{figure}

\begin{center}
{\bf 9. Conclusion}
\end{center}

The massive ideal leads which should be introduced for the
correct definition of the conductance of finite systems
have the essential influence on the properties of
the given sample.
In the present paper it was
demonstrated on the simplest example of 1D systems. In the
log-normal regime, this influence is reduced to the change of the
absolute scale of conductance, but generally changes the whole
distribution function. Under the change of the system length $L$,
the system resistance may undergo the periodic or aperiodic
oscillations.  Variation of the Fermi level induces the
qualitative changes in the conductance distribution, resembling
the smoothed Anderson transition.

The Economou--Soukoulis definition of conductance \cite{3}
refers to the composite system "sample+ideal leads", while
its relation to the system under consideration remains the
open question. For its solution, the introduction of
semi-transparent boundaries between the system and the ideal
leads was suggested in \cite{11}. In the limit of the weakly
transparent boundaries, one has the universal equations,
independent of the way how the contact resistance of the
reservoir is excluded \cite{10} (since all formulas of the
Landauer type \cite{2,5,6,8,9} reduces to the variant by
Economou--Soukoulis \cite{3,4}), which can be then
extrapolated to transparency of order unity.
Such definition refers surely to the given sample
and provides the infinite conductance for the ideal system.
However, the absolute scale of conductance is defined only in
the order of magnitude;
this point is not very significant
(only the ratio of conductances has a physical sense), but
creates certain difficulties in comparison with other results.
As clear from the present paper, the absolute scale of
conductance depends on the ideal leads, and in any case is a
conditional quantity. To avoid its uncertainty, one should
give information on the properties of leads: 
\onecolumn

\noindent
for example,
one can accept the "Gold Standard" and recalculate all results
to the leads made of gold.

According to Eq.2, the conductance distribution of an 1D
system is determined by two parameters  $\tilde\alpha L$
and $\gamma$. Such two parameter description is the natural
consequence of the one-parameter scaling hypothesis
\cite{38}, according to which the properties of a
$d$-dimensional cubical system of size $L$ are
completely determined by the ratio $L/\xi$, where $\xi$
is the correlation length. Composing the quasi-1D
system of size $L^{d-1}L_z$ from the cubical blocks of size
$L$, one comes to conclusion that its conductance depends on
the properties of one block ($L/\xi$) and a number of blocks
($L_z/L$).  For $L=a_0$ the quasi-1D system
becomes strictly one-dimensional, but is also
described by two parameters. One-parameter scaling in 1D systems
is realized in the deep of the allowed band  (when
equation (1) is valid) and approximately retains in the critical
region\,\footnote{\,In this case the parameter $\gamma\sim 1$
and does not have the essential evolution.}; its violation
in the forbidden band is rather natural due to violation of
the condition $\xi\gg a_0$. Necessity of the two-parameter
description of 1D systems was discussed in the paper \cite{34}
(see also \cite{21}--\cite{24})) and recently it was actively
used for description of the conductance distribution near the
Anderson transition \cite{12,35} in the framework of the Shapiro
scheme \cite{18}.

Evolution of the distribution function  $P(\rho)$, as
well as its moments, can be studied experimentally in the
spirit of the paper  \cite{37}, where different impurity
configurations in the given sample were created by its warming to
sufficiently high temperature.


\begin{center}
{\it Appendix 1.} {\bf Evolution of moments for the
coordinate transfer matrix} \end{center}

For the coordinate transfer matrix the following evolution
equations are valid
$$
z_{n+1}=(E-V_n) z_n +y_n\,,\qquad
y_{n+1}=- z_n \,,
\eqno(A.1)
$$
where $z_n=\tau^{(n-1)}_{12}$, $y_n=\tau^{(n-1)}_{11}$ with
the initial conditions $z_1=0$, $y_1=1$, or
$z_n=\tau^{(n-1)}_{22}$, $y_n=\tau^{(n-1)}_{21}$ with
the initial conditions $z_1=1$, $y_1=0$.
For the second moments one has
$$
\left ( \begin{array}{ccc} \overline{z^2_{n+1}} \\
\overline{z_{n+1} y_{n+1}} \\ \overline{ y^2_{n+1}}
\end{array} \right)\,
= \left ( \begin{array}{ccc} E^2+W^2 &  2E &\,\,\, 1\\
-E & -1 & \,\,\, 0 \\
1 & 0 &\,\,\, 0   \end{array} \right)\,
\left ( \begin{array}{ccc}
\overline{z^2_{n}} \\ \overline{z_{n} y_{n}} \\
\overline{ y^2_{n}}
 \end{array} \right) \,,
\eqno(A.2)
$$
and suggesting their exponential behavior
$\lambda^n$ with $\lambda=1+x$, obtains the
equation for
$x$
$$
x^3-x^2(E^2-4)-x(E^2-4)=W^2(2-3x-x^2)
\eqno(A.3)
$$
Setting $E^2-4=4\delta^2$, $W^2=4\epsilon^2\delta^2$ and taking
the limit $\delta\to 0$, $\epsilon\to 0$,
$\delta/\epsilon^2=const$, one can verify that (A.3)
coincides with (44).

Analogously, for the fourth moments one has a system of
equations
$$
\left ( \begin{array}{ccccc} \overline{z^4_{n+1} } \\
\overline{z^3_{n+1} y_{n+1}}
\\ \overline{z^2_{n+1} y^2_{n+1}} \\ \overline{z_{n+1} y^3_{n+1}}
\\ \overline{y^4_{n+1} } \end{array} \right)\,
= \left ( \begin{array}{ccccc} E^4+6E^2W^2 & \,\,4E^3+12EW^2  &
\,\,6E^2+6W^2 & \,\,4E &\,\,\, 1\\
-E^3-3EW^2 & -3E^2-3W^2 & -3E & -1 & \,\,\, 0\\
E^2+W^2 & 2E & 1 & 0 & \,\,\, 0\\
-E & -1 & 0 & 0 & \,\,\, 0\\
1 & 0 & 0 & 0 & \,\,\, 0
\end{array} \right)\,
\left ( \begin{array}{ccccc} \overline{z^4_{n} } \\
\overline{z^3_{n} y_{n}}
\\ \overline{z^2_{n} y^2_{n}} \\ \overline{z_{n} y^3_{n}}
\\ \overline{y^4_{n} } \end{array} \right)\,
\eqno(A.4)
$$
Suggesting that the moments behave as $\lambda^n$ with
$\lambda=1+x$, we have the equation for $x$
$$
-x^5+x^4(E^2+1)(E^2-4)+x^3(E^2-4)\left(-E^4+5E^2+1\right)
-x^2 2E^2(E^2-4)^2 -x E^2 (E^2-4)^2 +
$$
$$
+W^2\left[-6E^2(E^2-4) - 15x E^2(E^2-4) +x^2 (-12E^4+60E^2-6)
+\right.
$$
$$\left.
+x^3(-3E^4+30E^2-9) +x^4 (6E^2-3)\right] =0
\eqno(A.5)
$$
where terms of the higher order in $W^2$ are omitted. Setting
$E^2-4=4\delta^2$, $W^2=4\epsilon^2\delta^2$ and retaining the
main terms at the indicated limiting transition, we come
to Eq.48.

\twocolumn

\begin{center}
{\it Appendix 2.} {\bf The asymptotic forms of the
integral (70).} \end{center}

Using the evenness of the integrand of (70) in
variables  $\tilde \varphi=\varphi -\pi/2$, $\tilde \theta=\theta
-\pi/2$  and setting $x=\sin{\varphi}$, $y=\sin{\theta}$,
we have
$$
P(\rho)=\frac{1}{\pi^2\,t} \int\limits^{1}_{-1}\!
\frac{d x}{\sqrt{1-x^2}}
\int\limits^{1}_{-1}\! \frac{d y}{\sqrt{1-y^2}}
\exp\left\{-\frac{S(x,y)}{t}\right\},
\eqno(A.6)
$$
$$
S(x,y) =\left( \Delta_1 \sqrt{1\!+\!\rho}\, x +
\Delta_2 \sqrt{\rho}\, y  \right)^2 +\rho(1\!-\!y^2)  \,.
$$
Configuration of saddle points is essentially
different for $\rho<\rho_c$ and $\rho>\rho_c$, where
$\rho_c=\Delta_1^2$. In the first case the maximum of the
exponent is reached at $x=x_c$, $y=-1$ or $x=-x_c$, $y=1$,
where $x_c=\Delta_2 \sqrt{\rho}/\Delta_1 \sqrt{1\!+\!\rho}$,
while in the second case at $x=1$, $y=-1$ or $x=-1$, $y=1$.
Let $\delta x$ and $\delta y$ are deviations of $x$ and $y$
from the extremum point. For $\rho<\rho_c$ we retain in $S(x,y)$
the  quadratic term in $\delta x$ and the linear term in $\delta
y$, setting $x=x_c$, $1-y^2=2\delta y$ in the pre-exponential;
then
$$
P(\rho)=\frac{1}{\pi} \sqrt{\frac{1}{\rho(\rho_c\!-\!\rho)}} \,,
\eqno(A.7)
$$
which is the limiting form of the distribution for $t\to 0$.

For $\rho>\rho_c$ we have
$$
S(x,y) = S_c+A\delta x+ B\delta y +C\left(\delta x -x_c \delta
y\right)^2 -\rho (\delta y)^2\,,
\eqno(A.8)
$$
where $S_c$, $A$, $B$ are defined in Eq.73 and
$C=\Delta_1^2 (1\!+\!\rho)$. For large $\rho$ the linear terms
in $\delta x$, $\delta y$ are sufficient in
$S(x,y)$ and in radicals; then
$$
P(\rho)=\frac{1}{\pi} \sqrt{\frac{1}{AB}}
\exp\left(-\frac{S_c}{t}\right)\,,
\eqno(A.9)
$$
which is the last asymptotics in (72), (74)--(77).

The results $(A.7)$, $(A.9)$ are not applicable for $\rho$
close to $\rho_c$, since the coefficient $A$ turns to
zero at $\rho=\rho_c$. Setting $\delta \tilde x= \delta
x-x_c\delta y$ and omitting the last term in $(A.8)$, we have
$$
S(x,y) = S_c+A\delta \tilde x +
C\left(\delta \tilde x \right)^2 + 2\rho\delta y \,.
\eqno(A.10)
$$
For $A^2\ll Ct$ fluctuations of  $\delta\tilde x$ are determined
by the quadratic term, and $A\delta\tilde x$ can be omitted;
then in the small vicinity of $\rho_c$
$$
P(\rho)=\frac{1}{2\pi} \sqrt{ \frac{1}{2\pi\rho} }\,
 \frac{\Gamma(1/4)}{(4Ct)^{1/4}}
 \,, \qquad |\rho\!-\!\rho_c| \alt\,
\left(Ct\right)^{1/2} \,,
\eqno(A.11)
$$
so divergency at $\rho\to\rho_c$ is eliminated and the
third asymptotics in (72) and (75) is recovered. Under the
indicated condition, this result remains valid
for $\rho<\rho_c$, when $A$ is negative, but small in modulus. In
fact, Eq.$(A.11)$ takes place in the cases
$t\ll\Delta_1^2\ll 1$ and $t\ll 1\ll \Delta_1^2$, when the
fluctuation $\delta y \sim t/\rho$ is small in comparison with
$\delta\tilde x \sim \sqrt{t/C}$, so
$\delta\tilde x \approx \delta x$ and one can set $1-x^2=
2\delta\tilde x$, $1-y^2= 2\delta y$ in the pre-exponential.
 The inverse situation is realized in the case
$\Delta_1^2\gg 1$, $t\gg 1$,  when $\delta y\gg \delta\tilde x$
and $1-x^2\approx 2\delta x= 2\delta\tilde x +2x_c\delta y \approx
2\delta y$ in the pre-exponential; then we come to the result
$$
P(\rho)=\frac{1}{2\pi^2} \sqrt{\frac{\pi}{\rho\Delta^2_1 t}}
\ln \frac{\Delta^2_1 t}{\rho}
\exp\left(-\frac{S_c}{t} \right)  \,,
$$
$$
\qquad\qquad\qquad\qquad \rho_c/t \alt\rho\alt \rho_c t\,,
\eqno(A.12)
$$
determining the penultimate asymptotics in (76), (77). The
condition $A^2\ll Ct$ corresponds to $(\rho-\rho_c)^2\ll
\Delta_1^2\rho t$, which  for $t\gg 1$ is reduced to the given in
$(A.12)$. If $(A.11)$ is valid in the small vicinity of
$\rho_c$, the wide range of validity arises for $(A.12)$.

For small $\rho$ the saddle point approximation is not
applicable, and the initial form of (70) is
convenient.  If $t\ll \Delta_1^2$, then saddle point integration
over $\varphi$ is still possible and leads to expression
$$
P(\rho)=\frac{1}{\pi^2\,t} \int\limits^{\pi/2}_{-\pi/2} d\theta
\sqrt{ \frac{\pi t}{\Delta_1^2
   -\rho+\Delta_2^2\rho\,\sin^2{\theta}} } \,\cdot
$$
$$\,\,\quad\qquad\qquad\qquad\cdot\,
\exp\left\{-\frac{\rho}{t}\, \sin^2{\theta}\right\}
\,.
\eqno(A.13)
$$
For $\Delta_1^2\ll 1$ one has $\Delta_2\approx 1$; if
$\rho\ll t$, then only $\Delta_1^2$ can be retained in the
denominator, while integration over $\theta$ can be
produced by expansion of the exponent
$$
P(\rho)= \frac{1}{\pi} \sqrt{\frac{\pi}{\Delta^2_1 t}}
\exp\left(-\frac{\rho}{2t} \right) \,,\qquad  \rho\ll t\,,
\eqno(A.14)
$$
which is the first result in (72), (75), (76). For
$\rho\agt t$ one returns to the saddle point results
 $(A.7)$, $(A.9)$, $(A.10)$.

In the case  $\Delta_1^2\ll 1$ and $t\gg \Delta_1^2$ the
region of small $\rho$ is determined by the condition
$\rho(1+\rho)\ll t^2/\Delta_1^2$ and the integral (70)
is calculated by expansion over $S(\rho,\varphi,\theta)/t$ till
the second order, which gives the first result in  (74), (77);
for the opposite inequality we have
$A\approx B\approx 2\Delta_1 \Delta_2 \sqrt{\rho(1+\rho)}\gg t$
and $A^2\gg Ct$, which is sufficient for validity of
$(A.9)$.

In the case $\Delta_1^2\gg 1$ and  $t\ll \Delta_1^2$
Eq.$(A.13)$ remains valid, but its analysis is more
complicated. For $t\ll 1$ we have the familiar situation:
in the interval $\rho\alt t$ one can retain $\Delta_1^2$ in the
denominator and obtain $(A.14)$, while in the interval
$\rho\agt t$ the saddle point results $(A.7)$, $(A.9)$, $(A.10)$
are valid. For  $1\ll t\ll \Delta_1^2$ the result $(A.14)$
is valid only for $\rho\alt 1$. In the interval
$1\alt \rho\alt t$ the term $\Delta_2^2\rho\sin^2{\theta}$
is dominated in the denominator of $(A.13)$, while the
quantity $\Delta_1^2-\rho$ is necessary only for cutoff
of the logarithmic divergency:
$$
P(\rho)=\frac{1}{\pi^2} \sqrt{\frac{\pi}{\rho\Delta^2_1 t}}
\ln \frac{\Delta^2_1 \rho}{\rho_c-\rho}\,,
\qquad 1 \alt\rho\alt t\,.
\eqno(A.15)
$$
In the interval  $t\alt \rho\alt\rho_c$ the exponent
restrict integration in $(A.13)$ by values
$\theta^2\alt t/\rho$, so $\Delta_1^2 \rho$ in  $(A.15)$
is changed by $\Delta_1^2 t$. In fact, both results are actual
only for $\rho\ll\rho_c$, since in the interval
$\rho\agt \rho_c/t$ divergency at $\rho\to\rho_c$ is
eliminated due to nonlinear terms in $(A.8)$ and the
result $(A.12)$ is valid; so we have $\ln\rho$
for $1\alt\rho\alt t$ and $\ln{t}$ for $t\alt\rho\alt
\rho_c/t$, as is reflected in (76).

In the case $\Delta_1^2\gg 1$ and  $t\gg \Delta_1^2$,
expansion over  $S(\rho,\varphi,\theta)/t$ is possible in the
interval $\rho\alt t/\Delta_1^2$ and leads to the first result
(77). In the interval $\rho\agt t/\Delta_1^2$ expression
$(A.13)$ is valid, where $|\Delta_1^2-\rho|\ll t$,
$\Delta_2^2 \rho\gg t$ and the latter term is dominant in the
denominator; the logarithmic divergency is removed
due to restriction $(\Delta_2^2 \rho/t)\sin^2{\theta}\agt 1$,
which is necessary for the saddle point integration over
$\varphi$ and validity of $(A.13)$. If $\rho\alt t$,
then the exponent in $(A.13)$ is not essential and
the second result (77) holds. If $\rho\agt t$, then
we have the saddle point situation and validity of
$(A.12)$ and $(A.9)$.

\end{document}